\RequirePackage{fix-cm}
\documentclass[reqno,psamsfonts]{amsproc}
\usepackage{fixltx2e}
    \numberwithin{equation}{section}
\usepackage{amssymb}
\usepackage[citation-order,nobysame]{amsrefs}
\usepackage{pstricks,pst-plot}
\usepackage{hyperref}

\newcommand\rmd{\mathrm{d}}
\newcommand\rme{\mathrm{e}}
\newcommand\rmi{\mathrm{i}}
\newcommand\wt{\widetilde}

\newcommand\NW{\mathrm{NW}}
\newcommand\NE{\mathrm{NE}}
\newcommand\SE{\mathrm{SE}}
\newcommand\SW{\mathrm{SW}}

\renewcommand\leq{\leqslant}
\renewcommand\geq{\geqslant}

\newcommand\arctic{\mathcal{A}}
\newcommand\FO{\mathsf{F}}
\newcommand\DI{\mathsf{D}}

\newcommand\Gf{\mathcal{G}}

\newcommand\args[2]{\left(\genfrac{}{}{0pt}{}{#1}{#2}\right)}

\begin{document}

\title{The arctic curve of the domain-wall six-vertex model}

\author{F. Colomo}
\address{INFN, Sezione di Firenze\\
Via G. Sansone 1, 50019 Sesto Fiorentino (FI), Italy}
\email{colomo@fi.infn.it}

\author{A.G. Pronko}
\address{Saint Petersburg Department of V.A.~Steklov Mathematical
Institute of Russian Academy of Sciences\\
Fontanka 27, 191023 Saint Petersburg, Russia}
\email{agp@pdmi.ras.ru}

\begin{abstract}

The problem of the form of the `arctic' curve of the six-vertex model
with domain wall boundary conditions in its disordered regime is
addressed. It is well-known that in the scaling limit the model exhibits
phase-separation, with regions of order and disorder sharply separated by
a smooth curve, called the arctic curve. To find this curve, we
study a multiple integral representation for the emptiness formation
probability, a correlation function devised to detect spatial transition
from order to  disorder. We conjecture that the arctic curve, for
arbitrary choice of the vertex  weights, can be characterized by the
condition of condensation of almost all roots of the corresponding
saddle-point equations at the same, known, value.
In  explicit calculations we restrict to the disordered regime for which
we have been able to compute the scaling limit of certain generating function
entering the saddle-point equations.
The arctic curve is obtained in  parametric form and appears to be
a non-algebraic curve in general; it turns into an algebraic one in the so-called
root-of-unity cases. The
arctic curve is also discussed in application to the
limit shape of $q$-enumerated (with $0<q\leq 4$) large
alternating sign matrices. In particular, as $q\to 0$ the
limit shape tends to a nontrivial limiting curve, given
by a relatively simple equation.

\end{abstract}

\maketitle

\section{Introduction}

In strongly correlated systems, the effect of boundary conditions can be
relevant even in the thermodynamic limit. Consider for example a system whose
parameters are tuned in such a way that at equilibrium it should be in a
disordered phase, while its boundary conditions are chosen so that only
ordered configurations are admissible in the proximity of the boundary.
Due to the presence of strong correlations, it may happen that such boundary
conditions induce ordered regions extending macroscopically from the
boundaries deeply inside the bulk of the system. In such situation, spatial
phase separation emerges, with ordered regions contiguous to the boundary,
sharply separated from a central disordered region by a smooth curve,
called arctic curve in the case of dimer models \cite{JPS-98}.

Essentially the same phenomena appear in other contexts, with
different names, such as limit shape (in the statistics of Young diagrams
\cite{KV-77} and rhombi tilings \cite{CLP-98}), or interface (in
random growth models for two-dimensional crystals \cite{F-84}).
More generally, the problem consists in finding
the limit shapes and fluctuations of random two-dimensional surfaces
arising, for instance, in plane partitions
(or in three-dimensional Young diagrams, or in the melting of a faceted crystal)
\cite{CK-01,OR-01,FS-03},
and also in dimer models on planar bipartite graphs with fixed boundary conditions,
when described in terms of the height function \cite{KO-05,KOS-06}. For
recent developments see, e.g., \cite{E-09,BGR-09}. In these contexts, the
arctic curve is usually referred to as the frozen boundary of the limit
shape.

The standard example of an arctic curve is the famous arctic circle which
appeared in the study of domino tilings
of large Aztec diamonds \cite{EKLP-92,JPS-98}. The name originates
from the fact that in most  configurations the dominoes are
`frozen' outside the circle inscribed into the diamond, while
the interior of  the circle is a disordered, or `temperate',  zone. Further
investigations of the domino tilings of Aztec diamonds, such as details
of statistics near the circle, were also performed \cite{CEP-96,J-02,J-05}.

In the present paper we address the problem of the form of the arctic
curve in the six-vertex model with domain wall boundary conditions. The
case of generic Boltzmann weights of the disordered regime is considered.
We find an explicit expression for the curve in a parametric form; in general,
the curve appears to be  non-algebraic. This property  can be ascribed
to the fact that the six-vertex model cannot be reduced to a
model of discrete free (or Gaussian) fermions.

Indeed, in  all examples considered to date,  the arctic curves appear to
be algebraic curves. At the same time, these examples
can be seen as particular realizations of models of
discrete free fermions, although on various types of lattice and
with nontrivial boundary conditions.
Some  of them can be even reformulated as the six-vertex model at its
free-fermion point  with suitably chosen fixed boundary
conditions (the correspondence being however usually not bijective).
In particular, this is the case of domino tilings of Aztec
diamonds, and the corresponding boundary conditions of the
six-vertex model are exactly the domain wall ones
\cite{EKLP-92,FS-06}.
Thus the  problem of the arctic curve extends naturally  to
the six-vertex model with generic weights, and with fixed (in particular,
domain wall) boundary conditions.
The phenomenon of phase separation
for six-vertex model, in various regimes and with various fixed boundary
conditions, was studied
previously mainly numerically \cite{E-99,SZ-04,AR-05}; some analytical approaches
to treat the problem were discussed in \cite{Zj-02,PR-07}.

Historically, the six-vertex model with domain wall boundary conditions
was introduced to prove the Gaudin hypothesis for norms of Bethe states \cite{K-82}.
The standard framework for the model is the quantum
inverse scattering method, invented in seminal paper \cite{TF-79};
for a review and applications of the method see book \cite{KBI-93}.
The partition function of the model
was obtained in terms of a determinant, known as Izergin-Korepin
formula \cite{I-87,ICK-92}. The free energy per site was derived in
\cite{KZj-00,Zj-00}, where  phase separation was proposed
as a possible explanation for the observed influence of boundary conditions on
the thermodynamic properties. This, in turn, stimulated subsequent exact
calculation of correlation functions of the domain-wall six-vertex model
for generic values of its weights. This was done, using the quantum
inverse scattering method, for one-  and two-point
boundary correlation functions \cite{BPZ-02,CP-05c}, and, more recently,
also for a particular non-local correlation function, the
so-called emptiness formation probability \cite{CP-07b}.

The interest in the domain-wall six-vertex model is also motivated by its
close relationship with some problems in algebraic combinatorics. Apart from
the already mentioned domino tilings of Aztec diamond, the model was also
found to be related with enumerations of alternating sign matrices
\cite{EKLP-92}. This observation was useful since it opened new
possibilities in proving long-standing conjectures in this subject
\cite{Ku-96,Ze-96}; see also book \cite{Br-99} for a review. In this
context the arctic curve of the domain-wall six-vertex model is of great
interest since it describes the limit shape of very large alternating sign
matrices \cite{P-01,CP-08}.

In the present paper, to address the problem of the arctic curve, we
develop the idea proposed in \cite{CP-07b} of studying the scaling limit
of the emptiness formation probability,  a correlation function with the
capability of detecting spatial transition from order to disorder.
Namely, in the scaling limit the emptiness formation probability  may
have  only a simple step-function behaviour, when varying coordinates
from ordered to disordered regions, with the jump from one to zero
occurring exactly  at the arctic curve (actually,  at one of its four
portions, see Sections 2.3 and 4.1 for details). This can be seen as a
particular case  of the  general statement that the probability of
finding a macroscopically large ordered sub-region must vanish in the
disordered region. The arctic curve can then be obtained from certain
multiple integral representation  for the emptiness formation
probability, as the condition  on the relevant parameters (scaled
coordinates)  ensuring such a stepwise behaviour in the scaling limit.

This programme was first fulfilled in \cite{CP-07a} for a particular
choice of the parameter $\Delta$ of the six-vertex model, the case
$\Delta=0$, also known as the free-fermion point of the model. In that
paper it is observed that the arctic curve can be obtained by
investigating the multiple integral representation for emptiness
formation probability derived in \cite{CP-07b}. It was found that the
arctic curve corresponds to a rather peculiar solution of the system of
saddle-point equations, namely, the solution characterized by a trivial
Green function, with  just one pole. In other words, the arctic curve
appears to be in correspondence with the condition of `condensation' of
almost all saddle points at the same value in the complex plane. In the
case of $\Delta=0$ this correspondence, between the condensation and the
arctic curve,  admits a rigorous derivation since in this case the
saddle-point equations can be analysed using standard methods from the
theory of random matrix models (although some complication arises, due to
the two-cut nature of the problem, see Section 5.2, or also
\cite{CP-07a}, for details).

In the case of generic values of $\Delta$, i.e., away from the
free-fermion point, the system of coupled saddle-point equations
describing the scaling limit of the emptiness formation probability is
extremely intricate and the standard tools of random matrix models (e.g.,
the Green function approach) are inapplicable. We conjecture that the
arctic curve can nevertheless be found as the condition of condensation
of almost all roots of the saddle-point equations to the same, known,
value. As already observed in \cite{CP-08}, certain specific properties
of the investigated multiple integral representation, guaranteeing the
step-function behaviour of the emptiness formation probability in the
scaling limit, appear to be totally independent of the values of the
parameters of the model, in particular, of the value of $\Delta$. We also
propose a method for deriving the arctic curve without a direct use of
the Green function, but rather exploiting  the fact that in the case of
condensation the system of coupled saddle-point equations simplifies to a
single equation (that we call `reduced saddle-point equation'). This
equation must necessarily have two coinciding real roots, for consistency
with condensation itself. The arctic curve is then given as the condition
of coincidence of two roots of the reduced saddle-point equation, where
the value of this double root is the parameter which parameterizes
the curve.

To obtain the explicit form of the reduced saddle-point equation, and
hence to derive the arctic curve, one has also to find the thermodynamic
limit of certain function,  which plays the role of generating function
of a particular one-point boundary correlation function. Here we evaluate
this limit for the whole disordered regime, thus extending the results of
\cite{CP-08} where two relevant particular cases away from the
free-fermion point were considered. We also give here a detailed
exposition of the method.

The  organization  of the present paper mainly follows the logic of our
derivation of the main result. After recalling the definition of the
model, Izergin-Korepin formula, and the phase separation in Section 2, in
Section 3 we introduce boundary correlation functions and address the
problem of finding the contact points of the arctic curve with the
boundaries, in the disordered regime. The definition of the emptiness
formation probability, the multiple integral representation for this
correlation function, and the corresponding scaling limit saddle-point
equations are given in Section 4. We devote Section 5 to a derivation of
the arctic curve at the free-fermion point of the model. Here we also
explain how to derive the arctic curve without any use of the Green
function but through the simpler condition of coincidence of two roots of
the reduced saddle-point equation. In Section 6 we show that the  two
specific properties of the multiple integral representation for the
emptiness formation probability  which are relevant for the
correspondence between the arctic curve and condensation are totally
independent of the value of $\Delta$. Here we also derive the reduced
saddle-point equation and obtain the arctic curve in the disordered
regime ($|\Delta|<1$); formulae \eqref{XYzeta}--\eqref{Yzeta} constitute
the main result of the present paper.

In Section 7, we discuss some particular cases of the main result, also
in connection  with application of the model to the problem of
$q$-enumeration of alternating sign matrices (with $0<q\leq 4$). In
particular, we show that the limit shape of  $q$-enumerated alternating
sign matrices has a non-trivial limit as $q\to 0$; with a suitable choice
of coordinates the corresponding limiting curve  is just the cosine
curve. In Section 8 we summarize  the results and provide some concluding
remarks.

The text is followed by two appendices. In Appendix A we recall some
results on the connection between the Izergin-Korepin formula and the
generating function of a particular one-point boundary correlation
function. In Appendix B we obtain, for  the disordered regime, the
thermodynamic limit of this generating function, which is essentially
used for the derivation of the main result.

\section{Six-vertex model, domain wall boundary conditions,
and phase separation}

\subsection{The model}

The six-vertex model is canonically formulated in terms of
arrows pointing along the edges of a square lattice. The arrows must obey
the `ice-rule': there are two arrows pointing away from,
and two arrows pointing into, each lattice vertex.

Equivalently, one can describe the states of the edges
in terms of variables taking two values, e.g., $0, 1$.
We shall use the convention that such a variable takes value $0$ ($1$)
if the arrow on the edge is pointing upward or right (downward or left).
Let us consider a single vertex of the square lattice and
let $\mu$, $\nu$, $\nu'$ and $\mu'$ be the variables attached to the
left, bottom, top, and right edges of the vertex respectively. Let
$w(\mu,\nu,\nu',\mu')$ denote the Boltzmann weight associated with
the vertex. In the six-vertex model
\begin{equation}\label{icerule}
w(\mu,\nu,\nu',\mu')=0\quad\text{if}\quad \mu+\nu\ne\mu'+\nu'.
\end{equation}
Imposing the arrow-reversal invariance of the Boltzmann weights, we set
\begin{equation}\label{abc}
\begin{split}
w(0,0,0,0)=w(1,1,1,1)&=a,\\
w(0,1,1,0)=w(1,0,0,1)&=b,\\
w(0,1,0,1)=w(1,0,1,0)&=c,
\end{split}
\end{equation}
where $a$, $b$, and $c$ are some real positive constants, referred often
below as `weights'.

The partition function of the six-vertex model can be defined as
\begin{equation}\label{partition}
Z=\sum_{\{\mu\},\{\nu\}} \prod_{i,j}
w(\mu_{i,j},\nu_{i,j},\nu_{i,j}',\mu_{i,j}').
\end{equation}
Here the product is taken over vertices of the lattice and
the sum is performed over values of edge variables; the variables are subject
to the linking conditions
\begin{equation}\label{link}
\mu_{i,j}'=\mu_{i-1,j},\qquad \nu_{i,j}'=\nu_{i,j-1}.
\end{equation}
To define completely the model, one has also to specify the boundary
conditions in \eqref{partition}. We shall consider the so-called domain
wall boundary conditions, defined below.

{}From the analysis of thermodynamic limit
of the model with periodic boundary conditions (see, e.g. \cite{B-82})
it is well known that the model has three physical regimes, or phases;
two of them are regimes of order and one is a regime
of disorder.
Let us introduce the parameter
\begin{equation}
\Delta=\frac{a^2+b^2-c^2}{2ab}\;.
\end{equation}
The case $\Delta>1$ is called ferroelectric regime. Depending on
whether $a>b$ or $a<b$,  ground state configurations are built
from vertices of weight $a$, or $b$; in these configurations all arrows
are ordered ferroelectrically.
The case $\Delta<-1$ is called
anti-ferroelectric  regime. In this case,  the
doubly-degenerated ground state is formed only by vertices of weight $c$, and
thus the directions of arrows alternate
along vertical and horizontal rows of the lattice. Finally, the case
$|\Delta|<1$ is called disordered  (or critical) regime. In this case
typical configurations contain vertices of all three weights;
there is no particular order of arrows in these configurations.

All over the paper we use the following parameterization for the weights:
\begin{equation}\label{param}
a=\sin(\lambda+\eta),\qquad
b=\sin(\lambda-\eta),\qquad
c=\sin 2\eta.
\end{equation}
We also have
\begin{equation}
\Delta=\cos2\eta.
\end{equation}
The parameter $\lambda$
has meaning of a rapidity variable and $\eta$ is the so-called
crossing parameter.

In the case of the disordered regime, i.e., when
$a$, $b$, and $c$ are such that $|\Delta|<1$, both $\lambda$ and $\eta$ are
real and
satisfy
\begin{equation}\label{disorder}
\eta\leq \lambda \leq \pi-\eta,\qquad
0< \eta<\frac{\pi}{2}.
\end{equation}
The other two regimes, the ferroelectric and anti-ferroelectric
ones, can be approached by analytic continuation
(modulo a purely imaginary common factor for each weight),
choosing $\lambda=\rmi\tilde\lambda$, $\lambda=\rmi\tilde\eta$ and
$\lambda=\pi/2+\rmi\tilde\lambda$, $\lambda=\pi/2+\rmi\tilde\eta$,
respectively,
where $\tilde\lambda$ and $\tilde\eta$ are real, and satisfy some
further restrictions to ensure positivity of the weights.

For later reference, let us briefly discuss some symmetries of the model.
First, we mention the so-called crossing symmetry. It is a symmetry of
the Boltzmann weight under reflection of the vertex with respect to the
vertical (horizontal) line, reversing the values of the edge variables
on the horizontal (vertical) edges, and exchanging the values $a
\leftrightarrow b$. For example, the reflection with respect to
the vertical line gives the crossing symmetry  relation
\begin{equation}\label{crossing}
w(\mu,\nu,\nu',\mu'|\lambda)= w(\bar \mu',\nu,\nu',\bar\mu|\pi-\lambda).
\end{equation}
Here the bar means that the value of the edge variable must be
reversed, i.e., $\bar\mu=1$ ($\bar\mu=0$) if $\mu=0$ ($\mu=1$).
Similarly, one can also consider the
reflection with respect to the horizontal line, that gives an analogous
relation.

Another useful, and in fact even simpler, property of the six-vertex
model Boltzmann weight is its invariance under reflection of the vertex
with respect to one or another diagonal,
\begin{equation}
w(\mu,\nu,\nu',\mu'|\lambda)= w(\nu,\mu,\mu',\nu'|\lambda)=
 w(\nu',\mu',\mu,\nu|\lambda).
\end{equation}
We shall refer to this property as the diagonal-reflection symmetry.

\subsection{Domain wall boundary conditions}

Consider the six-vertex model on a square lattice obtained by intersection
of an equal number of vertical and horizontal lines.
Domain wall boundary conditions correspond to fixing the  variables on
the external edges of the lattice in a specific way,
namely, with all arrows on horizontal edges pointing outward the lattice,
and all arrows on external vertical edges pointing inward the lattice.
Equivalently, these boundary conditions mean that all variables
attached to external edges on the top and on the left are set to $1$,
while those attached to external edges on the bottom and on the right are set
to $0$.

Denoting by $N$ the number of vertical, or, equivalently, horizontal
lines of the lattice on which the model with domain wall boundary conditions
is considered, we denote
by $Z_N$ the partition function of this model,
\begin{equation}\label{ZN-def}
Z_N=\sum_{\{\mu\},\{\nu\}} \prod_{i,j=1}^{N}
w(\mu_{i,j},\nu_{i,j},\nu_{i,j}',\mu_{i,j}')\Bigg|_{
\substack{\mu_{N,*}=\nu'_{*,1}=1\\ \nu_{*,N}=\mu'_{1,*}=0}}.
\end{equation}
Here the stars indicate that the subscripts must run over all possible
values,
e.g., $\mu_{N,*}=1$ means that $\mu_{N,1}=\mu_{N,2}\dots=\mu_{N,N}=1$.

In parametrization \eqref{param}, it can be shown that
the partition function admits
the following exact representation \cite{I-87}
\begin{equation}\label{ZN-Hankel}
Z_N=
\frac{\big[\sin(\lambda-\eta)\sin(\lambda+\eta)\big]^{N^2}}{
\prod_{n=1}^{N-1}(n!)^2}\,
D_N.
\end{equation}
Here $D_N=D_N(\lambda)$ stands for the H\"ankel determinant
\begin{equation}\label{DN}
D_N:=
\begin{vmatrix}
\varphi(\lambda) & \varphi'(\lambda) & \dots &\varphi^{(N-1)}(\lambda)\\
\varphi'(\lambda) & \varphi''(\lambda) & \dots &\varphi^{(N)}(\lambda)\\
\vdots & \vdots & \ddots& \vdots\\
\varphi^{(N-1)}(\lambda) & \varphi^{(N)}(\lambda) & \dots
&\varphi^{(2N-2)}(\lambda)
\end{vmatrix},
\end{equation}
where function $\varphi(\lambda)$ is given by
\begin{equation}\label{varphi}
\varphi(\lambda):=\frac{\sin2\eta}{\sin(\lambda-\eta)\sin(\lambda+\eta)}.
\end{equation}
Formula \eqref{ZN-Hankel} follows from the more general result,
known as Izergin-Korepin formula (see Appendix A, equation \eqref{IKformula}),
which is valid for the model with inhomogeneous weights \cite{I-87}.
The proof of Izergin-Korepin formula is essentially based on the quantum
inverse scattering method \cite{TF-79,KBI-93}; see also \cite{ICK-92} for
details.

Using formula \eqref{ZN-Hankel} one can study the partition function in
the thermodynamic limit. The quantity of interest is the free energy per
site $f$, defined as
\begin{equation}
f:=-\lim_{N\to\infty}\frac{\ln Z_N}{N^2}.
\end{equation}
It can be shown \cite{KZj-00,Zj-00,BF-05} that the free energy per site of
the model in the disordered regime is given by the expression
\begin{equation}\label{freeenergy}
f=-\ln \left(\frac{\alpha\sin(\lambda+\eta)\sin(\lambda-\eta)}
{\sin\alpha(\lambda-\eta)}\right),
\end{equation}
where we have used the notation
\begin{equation}\label{alpha}
\alpha:=\frac{\pi}{\pi-2\eta}.
\end{equation}
For a derivation of formula \eqref{freeenergy} see also Appendix B.

Formula \eqref{freeenergy} was obtained for the first time in
\cite{KZj-00}, where also the case of ferroelectric regime was considered;
the case of the anti-ferroelectric regime was studied in \cite{Zj-00}.
In these studies, the following observation have been made: the free energy
per site is greater in the case of the domain wall boundary conditions,
in comparison with the case of periodic boundary conditions, both for the
disordered and anti-ferroelectric regimes, but remains the same for the
ferroelectric regime,
\begin{equation}
f_\text{domain wall}>f_\text{periodic}, \qquad
\text{for}\quad\Delta < 1.
\end{equation}
This result can be explained by the fact that in these two regimes, admissible
configurations are significantly constrained by the domain wall boundary
conditions. Moreover, for large lattices, these effects are macroscopically
large, i.e., the model exhibits phase-separation phenomena.

\subsection{Phase separation and arctic curves}

The six-vertex model exhibits spatial separation of phases for a wide choice
of fixed boundary conditions \cite{E-99}. Roughly speaking, the effect is
related to the fact that ordered configurations on the boundary can induce,
through the ice-rule, a macroscopic order inside the lattice. Analytically,
as
already mentioned, a signal for presence of phase separation is a change of
the free energy per site, in comparison with the case of periodic boundary
conditions.

The notion of phase separation acquires a precise meaning in the scaling
limit. This is a thermodynamic limit which is to be treated as a continuum
limit, namely, in this limit the number of lines of the lattice (in each
direction) tends to infinity and the lattice spacing vanishes, while the
total
size of the system (in each direction) is kept fixed.

For the domain-wall six-vertex model we take $N\to\infty$, but the whole
lattice is scaled to a finite square, e.g., with sides of length $1$. Typical
configurations are constrained to have macroscopic regions  of ferroelectric
order near the boundary. More precisely, if the parameters of the model are
tuned to the ferroelectric regime then there is full compatibility between
boundary and bulk phase, no conflict arises, and the whole system is in the
ferroelectric order. If, however, the  parameters of the model are tuned to
the disordered or the anti-ferroelectric regimes, then there is a competition
with the ferroelectric ordering induced by the boundaries, and phase
separation emerges. The present understanding of the phase-separation
phenomenon in these two regimes is mostly due to numerics
\cite{SZ-04,AR-05};
theoretical considerations based on a variational principle approach can be
found in \cite{Zj-02,PR-07}. Below we describe the case of disordered
regime,
and mention briefly the situation in the anti-ferroelectric regime at the
end.

\begin{figure}
\centering
\input{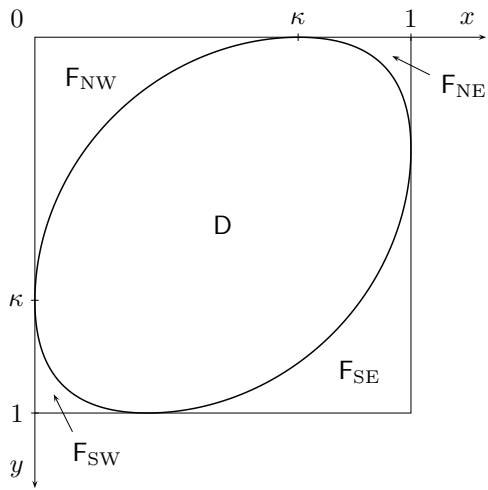}
\caption{The four regions of ferroelectric order and the region of disorder
for the model in the scaling limit, in the disordered regime.}
\label{fig-arctic}
\end{figure}

In the case of the disordered regime, in the scaling limit five regions
appear: four regions of ferroelectric order, $\FO_\NW, \FO_\NE, \FO_\SE,
\FO_\SW$ in the four corners of the square, and one region of disorder
$\DI$, in the centre, see figure \ref{fig-arctic}. The region of disorder
is sharply separated by some smooth curve  $\arctic$, called the arctic
curve. The arctic curve and the square have four contact points, located
each one on a side of the square.

The arctic curve in the disordered regime consists of four portions,
\begin{equation}
\arctic=\varGamma_\NW\cup\varGamma_\NE\cup\varGamma_\SE\cup\varGamma_\SW,
\end{equation}
where $\varGamma_i$ separates the region $\FO_i$  ($i=\NW,\NE,\SE,\SW$) from
the internal region of disorder $\DI$.

To describe, for example, the curve $\varGamma_\NW$, one can introduce a
function $\Upsilon(x,y;\lambda)$ where $x$ and $y$ are coordinates in the
scaling limit and $\lambda$ is the parameter of the weights,
\begin{equation}
\varGamma_\NW:\ \Upsilon(x,y;\lambda)=0,\qquad x,y\in[0,\kappa].
\end{equation}
It is to be emphasized, that the explicit form of $\Upsilon(x,y;\lambda)$
is significantly determined by the value of $\eta$. Given function
$\Upsilon(x,y;\lambda)$ one can readily obtain all other portions of the
arctic curve. Due to the crossing symmetry, relation \eqref{crossing}, we
have
\begin{equation}
\varGamma_\NE:\ \Upsilon(1-x,y;\pi-\lambda)=0,\qquad
x\in[1-\kappa,1], \ y\in[0,\kappa].
\end{equation}
Exploiting the diagonal-reflection symmetry, we further have
\begin{equation}
\varGamma_\SE:\ \Upsilon(1-x,1-y;\lambda)=0,\qquad x,y\in[1-\kappa,1].
\end{equation}
Finally, and essentially similarly, we have
\begin{equation}
\varGamma_\SW:\ \Upsilon(x,1-y;\pi-\lambda)=0,\qquad
x\in[0,\kappa], \ y\in[1-\kappa,1].
\end{equation}
Thus, knowledge of function $\Upsilon(x,y;\lambda)$ allows one to
determine the whole arctic curve $\arctic$.

It turns out that function $\Upsilon(x,y;\lambda)$ is algebraic (for
example, quadratic in $x$ and $y$) only for special values of $\eta$,
while in general it is a rather complicated non-algebraic, or
transcendental, function. To describe the curve $\varGamma_\NW$, in
practical calculation, we shall use the parametric form
\begin{equation}
x=X(\zeta),\qquad y=Y(\zeta),\qquad \zeta\in[0,\zeta_\text{0}],
\end{equation}
where $X(\zeta)$, $Y(\zeta)$ and $\zeta_\text{0}$ depend on the
parameters of the model, $\lambda$  and $\eta$. This must be taken into
account when reconstructing the whole arctic curve $\arctic$ as explained
above. Functions $X(\zeta)$ and $Y(\zeta)$ will be monotonously
decreasing and increasing functions, respectively, satisfying
$X(0)=\kappa$, $Y(0)=0$ and $X(\zeta_\text{0})=0$,
$Y(\zeta_\text{0})=\kappa$. Furthermore, these two functions are in fact
just a single function, since, due to reflection symmetry, we have the
relation $X(\zeta)=Y(\zeta_\text{0}-\zeta)$. The explicit results are
given in Section 6, and some interesting particular cases are discussed
in Section 7.

To conclude here, we mention briefly that in the case of the
anti-ferroelectric regime the picture of phase separation, though
reminding the situation of the disordered regime, is more complicated.
The competition between the anti-ferroelectric central region and  the
ferroelectric boundary regions induces the emergence of an intermediate
disordered region. As a result, there are two phase-separation curves: an
outer one, which is the usual arctic curve, separating the ferroelectric
and disordered phases, and an inner one separating the disordered and
anti-ferroelectric phases. We refer for further details to
\cite{SZ-04,AR-05} where numerical simulations for this regime were
performed.

\section{Boundary correlation functions}

\subsection{Definitions and properties}

Before addressing the whole problem of the arctic curve, it is useful to
restrict first to the  simpler problem of the position of the contact points.
This problem can be treated by studying correlation functions describing
probabilities of certain configurations near the boundaries, the so-called
boundary correlation functions.

Following \cite{BPZ-02}, we consider two kinds of boundary
correlation functions, denoted as $H_N^{(r)}$ and $G_N^{(r)}$
(in these notations the superscript in parenthesis, $r$, refers to some
distance on the lattice,
and should not be confused with an $r$th derivative with respect to any
variable).
As it will become clear later, the first correlation function
will play a somewhat fundamental role,
through its generating function $h_N(z)$ which enters various
correlation functions, while the
second one is more specific but relevant for addressing the problem of
the contact points.

The first correlation function, $H_N^{(r)}$, is designed to reflect
the fact that the configurations of the model with
domain wall boundary conditions admit  one and only one $c$-weight vertex
on each of the four boundary lines of the lattice.
Specifically, we define $H_N^{(r)}$ as the
probability that the sole $c$-weight vertex of the first
horizontal line from the top occurs at the $r$th position from the right.
It can be introduced by the formula
\begin{equation}\label{HNr-def}
H_N^{(r)}=
Z_N^{-1}\sum_{\{\mu\},\{\nu\}} \delta(\mu_{r,1},1) \delta(\mu'_{r,1},0)
\prod_{i,j=1}^{N}
w(\mu_{i,j},\nu_{i,j},\nu_{i,j}',\mu_{i,j}')\Bigg|_{
\substack{\mu_{N,*}=\nu'_{*,1}=1\\ \nu_{*,N}=\mu'_{1,*}=0}},
\end{equation}
where $\delta(\mu,\mu')$ stands for the Kronecker symbol,
\begin{equation}
\delta(\mu,\mu'):=
\begin{cases} 1 & \text{if}\quad \mu=\mu' \\
0 & \text{if}\quad \mu\ne\mu'
\end{cases}.
\end{equation}
This correlation function, satisfies the sum-rule
\begin{equation}\label{sumrule}
\sum_{r=1}^{N} H_N^{(r)}=1,
\end{equation}
which simply  expresses the fact that the total probability of finding a
$c$-weight vertex anywhere  in the first horizontal is exactly one.

Some properties of correlation function $H_N^{(r)}$ can be obtained using
the symmetries of the model. In particular, crossing symmetry implies
\begin{equation}\label{HNcrossing}
H_N^{(r)}(\lambda)=H_N^{(N-r+1)}(\pi-\lambda).
\end{equation}
Using the diagonal-reflection symmetry, one finds that
function $H_N^{(r)}$ also gives, for example,
the probability that the sole $c$-weight vertex on the rightmost vertical
line occurs at $r$th position from the top.

The second correlation function, $G_N^{(r)}$, is defined as the
probability of finding a given state on a horizontal edge of the first
line, i.e., it is  a particular case of polarization. For $r=1,\dots,N-1$
we define this correlation as the probability of having the edge variable
fixed to one, on the edge connecting $r$th and $(r+1)$th vertices (from
the right) of the first horizontal line (from the top),
\begin{equation}\label{GNr-def}
G_N^{(r)}=
Z_N^{-1}\sum_{\{\mu\},\{\nu\}} \delta(\mu_{r,1},1)
\prod_{i,j=1}^{N}
w(\mu_{i,j},\nu_{i,j},\nu_{i,j}',\mu_{i,j}')\Bigg|_{
\substack{\mu_{N,*}=\nu'_{*,1}=1\\ \nu_{*,N}=\mu'_{1,*}=0}}.
\end{equation}
We can also set $G_N^{(N)}=1$ and $G_N^{(0)}=0$ since the edge variables
on the leftmost and the rightmost external edges are already fixed by the
boundary conditions to $\mu_{N,1}=1$ and $\mu'_{1,1}=0$.

As noticed in  \cite{BPZ-02}, correlation functions $H_N^{(r)}$ and
$G_N^{(r)}$ are in fact closely related. Indeed, since
$\delta(\mu_{r,1},1)+\delta(\mu_{r,1},0)=1$, formulae \eqref{HNr-def} and
\eqref{GNr-def} imply that $G_N^{(r)}=H_N^{(r)}+G_N^{(r-1)}$. Taking into
account that $G_N^{(0)}=0$, one can equivalently bring this relation into
the form
\begin{equation}\label{GsumH}
G_N^{(r)}=H_N^{(r)}+H_N^{(r-1)}+\dots+ H_N^{(1)}.
\end{equation}
Recalling relation \eqref{sumrule}, we obtain that \eqref{GsumH} implies
that $G_N^{(N)}=1$, in agreement with the definition of this correlation
function.

As we shall see below, in the scaling limit function $G_N^{(r)}$ turns into
a step function; the point where its value jumps from $0$ to $1$
is exactly the contact point of the arctic curve. Relation \eqref{GsumH}
and certain properties of function $H_N^{(r)}$ will allow us to find the
location
of this point.

\subsection{Function $h_N(z)$ and its large $N$ limit}

The correlation function $H_N^{(r)}$ can be equivalently considered
through its generating function
\begin{equation}\label{hn}
h_N(z):=\sum_{r=1}^N H_N^{(r)} z^{r-1}.
\end{equation}
Due to \eqref{sumrule}, the generating function satisfies the
normalization condition $h_N(1)=1$. For later referring, let us also
mention the property
\begin{equation}\label{hN0}
h_N(0)=H_N^{(1)}=\frac{c a^{2N-2}Z_{N-1}}{Z_N}.
\end{equation}
This property simply follows from the fact that the $c$-weight vertex
appearing in the corner induces domain wall boundary conditions on the
remaining $(N-1)$-by-$(N-1)$ sub-lattice
\cite{BKZ-02}. We have also the relation
\begin{equation}
h_N(z;\lambda)=z^{N-1} h_N(z^{-1};\pi-\lambda),
\end{equation}
which is a direct consequence of relation \eqref{HNcrossing}.

Given function $h_N(z)$, and taking into account relation \eqref{GsumH},
one can readily represent correlation function $G_N^{(r)}$ as follows
\begin{equation}\label{Gint}
G_N^{(r)} = - \frac{1}{2\pi\rmi} \oint_{C_0} \frac{h_N(z) }{(z-1)z^r}\,
\rmd z.
\end{equation}
Here $C_0$ is a simple closed counterclockwise-oriented contour
surrounding point $z=0$, and lying in its small vicinity. In the next
Section, a much less trivial example of  correlation function will be
given, where a multi-variable generalization of function $h_N(z)$ will
appear.

In \cite{BPZ-02}, where the boundary correlation functions were
introduced, it was shown that they can be expressed similarly to formula
\eqref{ZN-Hankel} for the partition function, but with properly modified
entries in the last column of the determinant. Here we rely on a similar
but actually different representation, which relates generating function
$h_N(z)$ with the partition function of the `partially' inhomogeneous
model with a single parameter of inhomogeneity. This representation is
discussed in detail in Appendix A, see equations \eqref{wtZNxi} and
\eqref{wtZNhN}.

The following result plays a crucial role for what follows.

In the disordered regime, i.e., with $\lambda$ and $\eta$ satisfying
\eqref{disorder}, the function $\ln h_N(z)$,
for $z$ real and non-negative, has the following $N\to\infty$ behaviour
\begin{equation}\label{hNlargeN}
\ln h_N(\gamma(\xi))=
N\ln
\left(\frac{\sin\alpha(\lambda-\eta)\sin(\xi+\lambda-\eta)\sin\alpha\xi}
{\alpha \sin(\lambda-\eta)\sin\alpha(\xi+\lambda-\eta)\sin\xi} \right)
+ o(N),
\end{equation}
where
\begin{equation}\label{gamma}
\gamma(\xi)=
\frac{\sin(\lambda+\eta)\sin(\xi+\lambda-\eta)}{\sin(\lambda-\eta)
\sin(\xi+\lambda+\eta)}
\end{equation}
and $\alpha$ has been defined in \eqref{alpha}.

The proof of \eqref{hNlargeN} is given in Appendix B.

\subsection{Contact points}

To find the contact points in the disordered regime, i.e., the value of
constant $\kappa$, we study correlation function $G_N^{(r)}$ in the
scaling limit. Let us define function $G(x)$ ($0\leq G(x)\leq 1$)
as the scaling limit of correlation function $G_N^{(r)}$,
\begin{equation}
G(x):=\lim_{r,N\to\infty} G_N^{(r)},\qquad x=\frac{N-r}{N}, \qquad x\in[0,1].
\end{equation}
In the scaling limit, the dominating configurations are such that
all edge variables for horizontal edges  of the first line are fixed to $1$
in the region $\FO_\NW$, and to $0$ in the region $\FO_\NE$.
Therefore, $G(x)$ must have  a
simple stepwise behaviour, with the jump occurring precisely at the contact point,
\begin{equation}
G(x)=\begin{cases}
1& \text{for}\quad 0\leq x <\kappa \\
0& \text{for}\quad \kappa<x\leq 1 \\
\end{cases}.
\end{equation}
Hence,  the value of $\kappa$ can be found by addressing the question of how
correlation function $G_N^{(r)}$ may exhibit a stepwise behaviour, as
$N,r\to\infty$.

To address this question, we consider integral representation \eqref{Gint},
and apply the saddle-point technique to analyze its  behaviour in the
scaling limit. Formally, the saddle-point equation reads
\begin{equation}\label{sd}
-\frac{1-x}{z}+
\left(\lim_{N\to\infty}\frac{\ln h_N(z)}{N}\right)'=0.
\end{equation}
In principle, to investigate the solution of the saddle-point equation,
one would need to know the last term in \eqref{sd} for $z$ everywhere
in the complex plane. However,  it turns out that in addressing the problem of
contact points  the complete saddle-point analysis  can be avoided,
once the mechanism at the origin of the stepwise behaviour
of function $G(x)$ is understood. Furthermore, formula \eqref{hNlargeN},
which  was derived under the assumption of $z$ being real and positive,
appears to be sufficient.

Indeed, let us first restrict to the case $\Delta=0$, where function
$h_N(z)$ is given by a simple explicit expression,
$h_N(z)=[\tfrac{1}{2}(1-\sin2\lambda)z+\tfrac{1}{2}(1+\sin2\lambda)]^{N-1}$,
see \cite{BPZ-02,CP-05b} for a proof. From this expression it follows
that the saddle-point equation has only one solution, which we denote by
$z_\text{s.p.}$. It can be seen that $z_\text{s.p.}$, which depends on
the value of $x\in[0,1]$, is always real and positive, with the
corresponding  steepest-descent contour perpendicular to the real axis.
When deforming the contour $C_0$ in \eqref{Gint} to the steepest-descent
contour, the position of the saddle-point $z_\text{s.p.}$ with respect to
the pole at $z=1$ of the integrand appears to be crucial. Namely, for
$z_\text{s.p.}<1$, the pole at $z=1$ can simply be ignored in the
deformation of the contour, and  the saddle-point evaluation of the
integral vanishes in the scaling limit. On the other hand, for
$z_\text{s.p.}>1$, in deforming the contour through the saddle-point, we
necessarily  cross the pole at $z=1$, thus picking a contribution equal
to minus the residue of the integrand at this pole. This mechanism is at
the origin of the stepwise behaviour of $G(x)$, with the step occurring
when $x$ is such that $z_\text{s.p.}=1$. This last condition thus
determines the position of the contact point.

When considering the problem for generic values of $\Delta$, we note that
$h_N(z)$, being a polynomial, is analytic, and the only pole to be taken
into account in the integrand of \eqref{Gint} is again the one at $z=1$,
with residue $-1$, due to the property $h_N(1)=1$. Moreover, it can be verified that
saddle-point equation \eqref{sd} has only one solution when $z$ is real
and positive, as assumed in formula \eqref{hNlargeN}.
The stepwise behaviour
of $G(x)$ can thus be ascribed  again to the fact that, for some
particular value  $x=\kappa\in[0,1]$,  this real and positive saddle-point occurs
at $z=1$. This immediately determines the value of the contact point,
\begin{equation}\label{kappa}
\kappa=1-\left(\lim_{N\to\infty}\frac{\ln h_N(z)}{N}\right)'\bigg|_{z=1}.
\end{equation}
It is apparent that, for the evaluation of the contact point, the
knowledge of the large $N$ behaviour of $\ln h_N(z)$ for $z$ real and
positive, formula \eqref{hNlargeN}, is sufficient.

From expression \eqref{kappa}, and asymptotic behaviour \eqref{hNlargeN},
in the case of generic value of $\Delta$ in the disorder regime, we
obtain the following explicit expression
\begin{equation}
\kappa=\frac{\alpha\cot\alpha(\lambda-\eta)-\cot(\lambda+\eta)}
{\cot(\lambda-\eta)-\cot(\lambda+\eta)}.
\end{equation}
Note that, under the replacement $\lambda\mapsto\pi-\lambda$, we have
$\kappa\mapsto 1-\kappa$, as expected, due to the crossing symmetry. In
particular, for $\lambda=\pi/2$ we have $\kappa=1/2$ for all values of
$\eta$ in the disordered regime.

\section{Emptiness formation probability}

\subsection{Definition and basic properties}

We want now to address the problem of finding the arctic curve  $\arctic$
for the domain-wall six-vertex model. The most direct way would be to
consider a one-point correlation function, such as  polarization of all
horizontal (or vertical) edges on the lattice, compute for it a suitable
representation, and finally investigate its behaviour in the scaling
limit. This would give access not only to the arctic curve
$\arctic$ but to the whole `limit shape' of the model, in its `height
function' formulation, in the sense of \cite{PR-07}. The main obstruction
for such strategy resides in the difficulty of computing explicitly such
one-point correlation function. If one restricts to the arctic curve,
however, it is possible to devise some other correlation function, maybe
rougher than polarization, but nevertheless refined enough to
discriminate the spatial transition between order and disorder, and
simpler to calculate.

Indeed, consider  the emptiness formation probability (EFP), namely,
the probability $F_N^{(r,s)}$ of having the edge variables fixed to $1$
on  the first $s$ horizontal edges, from the top of the lattice, and located
between the $r$th and $(r+1)$th vertical lines, from the right \cite{CP-07b}
\begin{multline}\label{EFPdef}
F_N^{(r,s)} =
Z_N^{-1}\sum_{\{\mu\},\{\nu\}} \delta(\mu_{r,1},1)\delta(\mu_{r,2},1)
\cdots  \delta(\mu_{r,s},1)
\\ \times
\prod_{i,j=1}^{N}
w(\mu_{i,j},\nu_{i,j},\nu_{i,j}',\mu_{i,j}')
\Bigg|_{
\substack{\mu_{N,*}=\nu'_{*,1}=1\\ \nu_{*,N}=\mu'_{1,*}=0}}.
\end{multline}
Evidently, EFP is just a particular  case of a general $s$-point
correlation function, but it enjoys specific properties, outlined below,
which are suitable to address the problem of the arctic curve of the
model in the disordered regime.

By construction, EFP as defined in \eqref{EFPdef} actually measures the
probability that the edge variables of all edges  in the  top-left
$(N-r)\times s$ sub-lattice are fixed to $1$. This follows directly from
the definition of the six-vertex model, see condition \eqref{icerule},
and the domain wall boundary conditions. In other words, EFP measures the
probability that the top-left $(N-r)\times s$  sub-lattice has
ferroelectric order,

To find the arctic curve, one must address the asymptotic behaviour of  EFP
in the scaling limit. Namely, we are interested in the limit
$N,r,s\to\infty$, while keeping the ratios fixed. We set
\begin{equation}\label{scalingvar}
x:=\frac{N-r}{N},\qquad y:=\frac{s}{N},\qquad
x, y\in[0,1]\,.
\end{equation}
In this limit, coordinates $x$ and $y$ will parameterize
the unit square to which the lattice is rescaled.
Correspondingly, EFP is expected
to approach some limiting function
\begin{equation}
F(x,y) := \lim_{N,r,s\to\infty}F_N^{(r,s)}.
\end{equation}

Since EFP measures the probability that the   top-left $(N-r)\times s$
sub-lattice
has ferroelectric order, in the scaling limit, $F(x,y)$ should tend to $1$
whenever $(x,y)\in\FO_\NW$. On the other hand, in the
disorder region, by definition, the number of edges breaking the
ferroelectric
order
is macroscopic. Thus, as soon as $(x,y)$ enters the disordered region $\DI$, function
$F(x,y)$ should immediately vanish. As a result, in
the disordered regime, the limiting function is just the step function,
\begin{equation}\label{stepwise}
F(x,y)=\begin{cases}
1 &\text{if}\quad x,y \in \FO_\NW \\
0 &\text{if}\quad x,y \in \DI\cup\FO_\NE\cup\FO_\SE \cup \FO_\SW
\end{cases},
\end{equation}
where we have used notations of figure \ref{fig-arctic}.

Hence, EFP can be used to find curve $\varGamma_\NW$, which separates
region $\FO_\NW$ from the remaining part of the unit square. We recall,
see Section 2.3, that knowledge of curve $\varGamma_\NW$ allows one to
recover other portions $\varGamma_\NE$, $\varGamma_\SE$, and
$\varGamma_\SW$, thus solving the problem of the arctic curve of the
model.

Curve $\varGamma_\NW$ can be found by analysing a suitable representation for EFP
in the scaling limit. Moreover, formula \eqref{stepwise} implies that in such an analysis,
similarly to the case of the contact point considered in Section 3.3,
understanding the mechanism at the origin of the stepwise
behaviour of EFP in the limit is sufficient for finding the curve.

\subsection{Multiple integral representations}

In the following we resort to the representation for the emptiness formulation probability
in terms of $s$-fold integral obtained in \cite{CP-07b}.
The calculation itself is based
on the quantum inverse scattering method \cite{TF-79,KBI-93} and it
represents further development of ideas of \cite{BPZ-02,CP-05a}
where boundary correlation functions were studied. In fact, for what
follows we need two equivalent multiple integral representations, which
differ only in their integrand, one being just the totally symmetrized
version of the other.

To give these representations we need to introduce some notations first.
For $s=1,\dots,N$, let us define functions $h_{N,s}(z_1,\dots,z_s)$,
where the second subscript refers to the number of arguments, by the formula
\begin{multline}\label{hNs-def}
h_{N,s}(z_1,\dots,z_s) :=\prod_{1\leq i<k \leq s}^{}\frac{1}{z_k-z_i}
\\ \times
\begin{vmatrix}
z_1^{s-1} h_{N-s+1}(z_1) & \hdots & z_s^{s-1} h_{N-s+1}(z_s)\\
z_1^{s-2}(z_1-1) h_{N-s+2}(z_1) & \hdots & z_s^{s-2}(z_s-1) h_{N-s+2}(z_s)\\
\vdots & \ddots &\vdots\\
(z_1-1)^{s-1} h_{N}(z_1) & \hdots & (z_s-1)^{s-1} h_{N}(z_s)
\end{vmatrix}.
\end{multline}
Note that we also have $h_{N,1}(z)=h_N(z)$.
The functions  $h_{N,s}(z_1,\dots,z_s)$ are symmetric
polynomials of degree $(N-1)$ in each variable $z_1,\dots,z_s$.
In fact, these functions can be recursively constructed
starting with function $h_{N,N}(z_1,\dots,z_N)$, since,
due to the structure of \eqref{hNs-def}, it is easy to observe the relation
\begin{equation}\label{hnszto1}
h_{N,s}(z_1,\dots,z_{s-1},1)=h_{N,s-1}(z_1,\dots,z_{s-1}).
\end{equation}
One more relation of the same kind is
\begin{equation}\label{hnszto0}
h_{N,s}(z_1,\dots,z_{s-1},0)=h_N(0) h_{N-1,s-1}(z_1,\dots,z_{s-1}).
\end{equation}
where the value of $h_{N}(0)$ is given by formula \eqref{hN0}.
As shown in \cite{CP-06,CP-07b},
functions $h_{N,s}(z_1,\dots,z_s)$ are closely related to
the partition function of the partially inhomogeneous six-vertex model
with domain wall boundary conditions, with $s$ nonzero inhomogeneities.
This relationship is sketched in Appendix A.

We are now ready to turn to the multiple integral representations. The first
multiple integral representation reads:
\begin{multline}\label{MIR1}
F_N^{(r,s)} = \frac{(-1)^s}{(2\pi \rmi)^s}
\oint_{C_0}^{} \cdots \oint_{C_0}^{}
\prod_{j=1}^{s}\frac{[(t^2-2\Delta t)z_j+1]^{s-j}}{z_j^r(z_j-1)^{s-j+1}}\,
\\ \times
\prod_{1\leq j<k \leq s}^{} \frac{z_j-z_k}{t^2z_jz_k-2\Delta t z_j+1}\;
h_{N,s}(z_1,\dots,z_s)
\,\rmd z_1\cdots \rmd z_s.
\end{multline}
Here $C_0$, as before, denotes a simple anticlockwise oriented contour
surrounding the point $z=0$ and no other singularity of the integrand.
We use, besides
the anisotropy parameter $\Delta$, the asymmetry parameter
\begin{equation}
t:=\frac{b}{a}.
\end{equation}
The second, essentially equivalent, representation reads:
\begin{multline}\label{MIR2}
F_N^{(r,s)}=\frac{(-1)^{s(s+1)/2}Z_s}{s!(2\pi\rmi)^s a^{s(s-1)}c^s}
\oint_{C_0}^{} \cdots \oint_{C_0}^{}
\prod_{j=1}^{s} \frac{[(t^2-2\Delta t)z_j+1]^{s-1}}{z_j^r(z_j-1)^s}\,
\\ \times
\prod_{\substack{j,k=1\\ j\ne k}}^{s} \frac{1}{t^2 z_jz_k-2\Delta t z_j +1}
\prod_{1\leq j<k\leq s}^{} (z_k-z_j)^2
\\ \times
h_{N,s}(z_1,\dots,z_s)h_{s,s}(u_1,\dots,u_s)
\,\rmd z_1\cdots \rmd z_s.
\end{multline}
Here $Z_s$ denotes the partition function of the six-vertex model
with domain wall boundary conditions on an $s\times s$ lattice, and
\begin{equation}\label{uofz}
u_j:=-\frac{z_j-1}{(t^2-2\Delta t)z_j+1}.
\end{equation}
The integrand of \eqref{MIR2} is just the symmetrized version of the
integrand of \eqref{MIR1}, with respect to permutations of the
integration variables $z_1,\dots,z_s$. This follows through certain
symmetrization procedure, see Appendix D of \cite{KMST-02}, and an
additional identity proven in \cite{CP-07b}.

Representations \eqref{MIR1} and \eqref{MIR2} hold for all the three
regimes of the model.

\subsection{Saddle-point equations}

We are interested in the behaviour of EFP in the so-called
scaling limit, that is in the limit where $r$, $s$ and  $N$  are all
large, with the ratios $r/N$ and $s/N$ kept finite (and smaller than $1$).
In this limit, in principle, EFP can be analysed through
the saddle-point method applied to multiple
integral representation \eqref{MIR2}.

To find the corresponding system of saddle-point equations, we apply
standard arguments (familiar, for example, from the matrix model context)
to the integrand of \eqref{MIR2}. We have to use the fact that quantities
like $\ln h_{N,s}(z_1,\dots,z_s)$ are of order $s^2$, and that their
derivatives with respect to $z_j$'s are of order $s$. As a result, we
arrive at the following system of coupled saddle-point equations
\begin{multline}\label{SPE}
-\frac{s}{z_j-1}
-\frac{r}{z_j}
+\frac{s(t^2-2\Delta t)}{(t^2-2\Delta t)z_j+1}
-\sum_{\substack{k=1\\ k\ne j}}^{s}
\bigg(\frac{ t^2 z_k-2\Delta t}{t^2z_jz_k-2\Delta t z_j +1}
\\
+\frac{t^2 z_k}{t^2z_jz_k-2\Delta t z_k +1}+\frac{2}{z_k-z_j}\bigg)
+ \frac{\partial\ln h_{N,s}(z_1,\dots,z_s)}{\partial z_j}
\\
-\frac{t^2-2\Delta t+1}{[(t^2-2\Delta t)z_j+1]^2}\;
\frac{\partial\ln h_{s,s}(u_1,\dots,u_s)}{\partial u_j}
=0,
\end{multline}
where $j=1,\dots,s$. In writing this expression we neglect all sub-leading
contributions (i.e., estimated as $o(s)$).

As we shall see, although the saddle-point analysis of the multiple
integral representation for EFP cannot be actually performed in the full
rigour in general, nevertheless we will be able to find the condition on
the ratios $r/N$ and $s/N$ which corresponds to the jump from $0$ to $1$
of EFP in the scaling limit, i.e., to derive the equation for the arctic
curve. In many respects, the whole procedure will remind the saddle-point
analysis of the single integral \eqref{Gint} which allowed us to find the
location of the contact point.

\section{An example: the case $\Delta=0$}

\subsection{Preliminaries}

Before to proceed with the general case, it is useful to
consider technically the simplest case  $\Delta=0$.
The approach described in this section
was given in detail in  \cite{CP-07a}. Our aim here
is to recall the main steps of the approach and formulate a modified version
of it, which appears sufficiently simple and powerful to be applied
to the much more complicated case of generic values of $\Delta$
in the disordered regime, $|\Delta|<1$.

In the case $\Delta=0$, or, equivalently, $\eta=\pi/4$,
function $h_N(z)$ is known explicitly. It has a very simple form
(see, e.g.,  \cite{CP-05b})
\begin{equation}
h_N(z)=\left(\frac{t^2z+1}{t^2+1}\right)^{N-1},\qquad t
=\tan\big(\lambda-\tfrac{\pi}{4}\big).
\end{equation}
The determinant entering the definition of functions $h_{N,s}(z_1,\dots,z_s)$
turns out to reduce
simply to a Vandermonde determinant; straightforward calculation gives us
\begin{equation}
h_{N,s}(z_1,\dots,z_s)=
\prod_{j=1}^{s} \left(\frac{t^2z_j+1}{t^2+1}\right)^{N-s}
\prod_{1\leq j<k\leq s}^{} \frac{t^2z_jz_k+1}{t^2+1}.
\end{equation}
From this formula, using $u_j=(1-z_j)/(t^2z_j+1)$, we also have
\begin{equation}
h_{s,s}(u_1,\dots,u_s)=
\prod_{j=1}^{s}\frac{1}{(t^2z_j+1)^{s-1}}\prod_{1\leq j<k\leq s}^{}
(t^2z_jz_k+1).
\end{equation}
Taking into account that in the considered case $Z_s=1$ (see, e.g., \cite{BPZ-02})
and hence $Z_s/a^{s(s-1)}c^s=(t^2+1)^{s(s-1)/2}$, we obtain, as a result, that
representation \eqref{MIR2} at $\Delta=0$ reads:
\begin{multline}\label{MIRforFF}
F_N^{(r,s)} = \frac{(-1)^{s(s+1)/2}}{s!(2\pi\rmi)^s (t^2+1)^{s(N-s)}}
\oint_{C_0} \cdots\oint_{C_0}
\prod_{j=1}^s
\frac{(t^2 z_j+1)^{N-s}}{ z_j^r(z_j-1)^s}
\\ \times
\prod_{1\leq j<k\leq s} (z_j-z_k)^2\,
\rmd z_1\cdots \rmd z_s.
\end{multline}
Due to the last factor of
the integrand, which is the squared Vandermonde determinant, formula
\eqref{MIRforFF} naturally recalls the random matrix model partition
function, that will be exploited in what follows.

Here we would like also to mention that the last expression can be brought
into the
form
\begin{equation}\label{detQ}
F_N^{(r,s)}=\det Q
\end{equation}
where matrix $Q$ is an $s$-by-$s$ matrix with the entries
\begin{equation}
Q_{jk}=-\frac{1}{2\pi\rmi (t^2+1)^{N-s}}\oint_{C_0}^{}
\frac{(t^2 z+1)^{N-s}}{z^r (z-1)^s}\,(z-\theta)^{j-k+s-1}\,\rmd z.
\end{equation}
Here $\theta$ is an arbitrary parameter whose value does not affect
the determinant in \eqref{detQ}.
Note that matrix $Q$ has the structure of a Toeplitz matrix,
i.e., its entries depend on the indices only through their difference.

The determinant formula \eqref{detQ} has the following important implication.
Let us consider the quantity
\begin{multline}
I_N^{(r,s)} := \frac{(-1)^{s(s+1)/2}}{s!(2 \pi\rmi)^s (t^2+1)^{s(N-s)}}
\oint_{C_1^-} \cdots\oint_{C_1^-}
\prod_{j=1}^s
\frac{(t^2 z_j+1)^{N-s}}{z_j^r (z_j-1)^s}
\\ \times
\prod_{1\leq j<k\leq s} (z_j-z_k)^2\,
\rmd z_1\cdots \rmd z_s.
\end{multline}
This quantity differs from \eqref{MIRforFF} only in the integration contours:
here $C_1^-$ is a closed clockwise oriented contour (the minus sign
in the superscript indicates negative direction) in the complex plane
enclosing point
$z=1$, and no other singularity of the integrand. Since the integrand is kept
intact, we
have
\begin{equation}\label{I=detS}
I_N^{(r,s)}=\det S,
\end{equation}
where entries of matrix $S$ differ from those of matrix $Q$
only in the integration contour,
\begin{equation}
S_{jk}=-\frac{1}{2\pi\rmi (t^2+1)^{N-s}}\oint_{C_1^-}^{}
\frac{(t^2 z+1)^{N-s}}{z^r(z-1)^s}\, (z-\theta)^{j-k+s-1}\,\rmd z.
\end{equation}
Here  $\theta$ is again an arbitrary parameter, whose value does not affect
the determinant in \eqref{I=detS}. Setting $\theta=1$, matrix $S$ reduces to
an
upper-triangular matrix, with all its
diagonal entries equal to $1$. Hence
\begin{equation}\label{I=1_FF}
I_N^{(r,s)}=1,
\end{equation}
for all $r,s=1,\dots,N$. Identity \eqref{I=1_FF}, together with some
other properties of representation \eqref{MIRforFF}, turns out to be one
of the main ingredients entering the derivation of the arctic curve.

\subsection{Matrix model approach}

We want now to address the asymptotic behaviour in the scaling limit,
as  defined in Section 4.1, of  multiple integral representation
\eqref{MIRforFF}  for EFP in the case $\Delta=0$.
Inspired by the obvious analogy with the large $s$ limit of the partition
function of an   $s\times s$  Random Matrix Models,
we rewrite multiple integral representation \eqref{MIRforFF} as follows:
\begin{multline}\label{MIRasRMM}
F_N^{(r,s)} =
\frac{(-1)^{s(s+1)/2}}{s! (1+t^2)^{(1/y-1)s^2}(2 \pi i)^s}
\oint_{C_0} \cdots\oint_{C_0}
\exp\Bigg\{- s \sum_{j=1}^s V(z_j)\Bigg\}
\\ \times
\prod_{1\leq j< k\leq s} (z_j-z_k)^2
\rmd z_1\cdots \rmd z_s
\end{multline}
where the potential is given by
\begin{equation}
V(z_j)=\ln(z_j-1)+\frac{1-x}{y}\ln z_j -\left(\frac{1}{y}-1\right)\ln(t^2 z_j+1).
\end{equation}
Here we have used the scaling variables $x$ and $y$ introduced in
\eqref{scalingvar}.
The corresponding system of coupled saddle-point equations reads
\begin{equation}\label{SPE_FF}
-\frac{1}{z_j-1}-
\frac{1-x}{y z_j}+\frac{(1-y)t^2}{y(t^2 z_j+1)} +
\frac{2}{s}\sum_{\substack{k=1\\ k\ne j}}^{s}\frac{1}{z_j-z_k}=0,
\end{equation}
where $j=1,\dots,s$.

The standard physical picture reinterprets the saddle-point equations
as the equilibrium conditions for the position of $s$  particles,
each with charge $1/s$,  with logarithmic electrostatic repulsion,
in an external potential.  In the present case
the latter is built from three logarithms, and can be seen as generated
by three external charges, $1$, $(1-x)/y$, and $-(1/y-1)$
at positions $1$, $0$, and $-1/t^2$, respectively.
It is natural to refer to this model as the triple Penner
model. Although the simple Penner \cite{P-88} matrix model
has been widely investigated, not so much is known about the more
complicate multiple or generalized Penner models.
The main ideas we shall use in the following
are based on the phenomenon of eigenvalues condensation, typical of
Penner  models, and nicely described in  \cite{PW-95,AMK-94}.

To investigate the structure of solutions of the saddle-point
equations \eqref{SPE_FF} for large $s$, one can start
with introducing  the Green function
\begin{equation}\label{green}
\Gf(z):= \frac{1}{s}\sum_{j=1}^s \frac{1}{z-z_j},
\end{equation}
which, if the $z_j$'s solves \eqref{SPE_FF},  has
to satisfy some particular differential equation, which can be derived
by standard means. In the scaling  limit, such differential equation
reduces to an algebraic one for the limiting Green function. In principle,
the latter can be solved, and an expression for $\Gf(z)$ can be obtained in the
large
$s$ limit. In this limit, the poles of $\Gf(z)$, organize into cuts in the
complex plane,
and the discontinuities across such cuts define, when real and positive,
the density of solutions of the saddle-point equation as $s\to\infty$.
In the present case, however, the Green function $\Gf(z)$ is not completely
determined by the above procedure, since it contains an extra parameter,
which can be fixed only implicitly. This is a direct consequence
of the `two-cut' nature of the problem (see \cite{CP-07a} for details).
The problem of explicitly finding the density of solutions of
saddle-point equation \eqref{SPE_FF} as $s\to\infty$, for generic  values
of $x$, $y$, is therefore a formidable one, not to
mention the evaluation of the corresponding `free energy', and of
the saddle-point contribution to the integral in \eqref{MIRasRMM}.

The problem we are addressing is fortunately much more modest:
we are presently interested only in the expression of curve
$\varGamma_\NW$, which, as explained in Section 4.1,
corresponds to the curve in the unit square, where, in the scaling  limit,
EFP has a unit jump.
As discussed in Section 3.3, concerning the scaling limit
of the  boundary correlation function   $G_N^{(r)}=F_N^{(r,1)}$,  its
stepwise
behaviour was related to the position of the saddle-point solution
with respect to the pole at $z=1$. It is easy to verify that
the same mechanism holds for $F_N^{(r,s)}$ for any finite value of $s$:
indeed, in the $r,N\to\infty$ limit, with
$s$ and $r/N=1-x$ kept fixed (and, obviously, $s/N=y\to 0$), saddle-point
equations \eqref{SPE_FF}
decouple into $s$ identical  saddle-point equations of the form
\eqref{sd}. The line of reasoning discussed in  Section 3.3 can be applied
again, and due also to identity  \eqref{I=1_FF}, it is clear that
the unit jump defining the position of the contact point occurs now
when all $s$ saddle-point  solutions are located at $z=1$.

Before going on to the large $s$ situation, it is worth to recall
a peculiar property of the saddle-point solutions for generic  Penner
models (see \cite{PW-95} and \cite{AMK-94} for  further details).
In these models the logarithmic wells in the potential
can behave as condensation germs for the saddle-point solutions. It can
be shown that condensation can occur only for charges less than or
equal to $1$, the value of the charge corresponding to the fraction
of condensed solutions. Within the electrostatic analogy, this corresponds
to the capability for the logarithmic potential well created by external
charge $Q$ to screen exactly a fraction $Q$ of the $s$ particles
of charge $1/s$.

In our case, the only possibility for condensation of solutions
of the saddle-point equations is given by  the charge at $z=1$ in
the triple Penner potential,
since the charge at $z=-1/t^2$ is always repulsive, while the one at
$z=0$ is larger than $1$ in the region where $\varGamma_\NW$ lies.
Note further that in our case the charge at $z=1$ is exactly $1$,
allowing for total condensation of solutions of the saddle-point equations.
This consideration, and the crucial identity \eqref{I=1_FF},
strongly suggests that the mechanism producing the stepwise behaviour
of  $F_N^{(r,s)}$ in the scaling limit is still the same as in the case
of finite $s$, and thus that the curve $\varGamma_\NW$ occurs
in correspondence to the condensation of (almost) all solutions of
saddle-point equations \eqref{SPE_FF} at the point $z=1$

In the framework of the Green function approach, our claim can be rephrased
as follows: the curve $\varGamma_\NW$ can be derived from
the condition that $x$ and $y$ should be such that in
the scaling limit Green function \eqref{green} reduces to
\begin{equation}\label{Greeneq2}
\Gf(z)=\frac{1}{z-1}.
\end{equation}
Indeed, in \cite{CP-07a} it is shown that this requirement
translates into  the following  condition:
\begin{equation}\label{arctic_ellipses}
(1+t^2)(x+y-1)^2 +t^2 (1+t^2) (x-y)^2=t^2,
\end{equation}
for $x,y\in[0,\kappa]$. Recalling how the remaining portions of the arctic
curve can be
reconstructed (see discussion in Section 2.3), we also find that
equation \eqref{arctic_ellipses} describes the whole curve $\arctic$, i.e.,
it is valid for all $x,y\in[0,1]$.
At $\lambda=\pi/2$, or $t=1$, it turns into
$ (x-1/2)^2+ (y-1/2)^2=1/4$, which is the arctic circle of \cite{JPS-98}.

\subsection{Condensation of roots and arctic curve}

As just discussed,
the arctic curve $\varGamma_\NW$ occurs in correspondence to the
situation where almost all roots of saddle-point equations
\eqref{SPE_FF} condense at point $z=1$. It turns out that
this observation can be used to formulate a simpler approach,
which does not involve
the Green function, and can be extended to more general situations
(beyond the case $\Delta=0$).

The origin of the observed correspondence between
the arctic curve $\varGamma_\NW$ and the condensation of almost
all roots of saddle-point equations relies
on the two following important ingredients.  On one hand,
by construction, EFP in the scaling limit has  a stepwise behaviour,
the unit jump occurring in correspondence to  $\varGamma_\NW$.
On the other hand, generalized Penner models (whose saddle-point
equations share some essential features with the ones under consideration)
allow for partial or total condensation of roots.
The role of the  unit charge potential in a generalized Penner model
with total condensation is played in our case by the $s$ poles
at $z=1$ in representation \eqref{MIRforFF}, which all turn out to be poles
of order $s$.

Having in mind these essential features,
the correspondence between the arctic curve and the condensation of roots
can be seen as the consequence of the following properties
of multiple integral representation for EFP:
\begin{itemize}
\item
in each integration variable $z_1,\dots,z_s$ the integrand has a  pole at
$z=1$, and the cumulative residue over all these variables
at this pole is equal to $1$;
\item the pole at $z=1$ for each integration variable $z_1,\dots,z_s$ is
of
order $s$.
\end{itemize}
The first property means that the scaling limit of EFP
is governed by the position of the roots of saddle-point equations
with respect to the pole at $z=1$, and, furthermore, identity \eqref{I=1_FF}
implies that
deforming the integration contours trough this pole produces exactly
the expected stepwise limiting expression for EFP.
The second property means that point $z=1$ is the point where
the condensation of almost all roots of the corresponding saddle-point
equations
may occur.

The properties listed above, together with the expected
limiting expression \eqref{stepwise}, implies that the arctic curve
can be found through the condition of condensation  of
almost all roots of the saddle-point  equations at $z=1$.
Let $n_c$ and $n_u$ denote the number of condensed and uncondensed roots,
respectively, $n_c+n_u=s$. Condensation of `almost all' roots,
simply means that
\begin{equation}\label{n_u}
\frac{n_c}{s}\sim 1, \qquad \frac{n_u}{s}\sim 0, \qquad s\to\infty.
\end{equation}
Note that condensation of almost all roots at $z=1$ evidently implies formula
\eqref{Greeneq2} and also that RHS of \eqref{SPE_FF}
reduces to $2/(z_j-1)$, at leading order for large $s$.
Hence, the system of saddle-point equations simplifies to a system
of $n_u$ identical, decoupled, equations.  This is in fact one single
equation determining the position of the $n_u$ uncondensed roots.
This equation can be called `reduced saddle-point equation'.
It turns out that the solutions of the
reduced saddle-point equation, due to condensation itself,
must obey certain property.

According to the standard picture, described in  \cite{AMK-94},
the solutions of the saddle-point equations build
up cuts in the large $s$ limit, and these cuts move in the complex
plane, as the parameters of the model
(in our case $x$ and $y$) are varied.
Specifically, in our case, when $(x,y)\in\DI$ (see figure \ref{fig-arctic}),
there is a  cut, with complex conjugate end-points (let them
be $z_a$ and $z_b$), which lies to the left of the singularity at $z=1$,
and intersects the real axis on the segment $(0,1)$.
When we move  $(x,y)$ closer to   the frozen region, $\FO_\NW$,
the end-points of the cut, while still complex conjugated, get closer to
the real axis, with real part greater than $1$.
When $(x,y)$ reaches  the arctic curve, the end-points of the cut join
at some value $w$ on the real axis, $w>1$, with the cut entangled around
the singularity at $z=1$. This corresponds to the total condensation
of solutions of the saddle-point equations. When $(x,y)$ enters the frozen
region $\FO_\NW$, the end-points separates,  still sticking to the real axis,
generating  now a cut to the right of the singularity at $z=1$.

Concentrating on the case of total condensation, the cut goes now
from $z_a=w$ just above the real axis to the singularity at $z=1$,
surrounds it and comes back to $z_b=w$, running just below the real axis.
The two portions of the cut which lie one in front of the other just
above and below the real axis somehow compensate (the
corresponding densities of roots cancel each other), and
one is left with a pole at $z=1$, describing the total condensation of roots.
Hence, in this case, among the uncondensed roots, it is necessary to
have a pair of coinciding roots, with value $w$,
corresponding in fact to the end-points of the would-be cut, now collapsed
to the pole at $z=1$. Such pair of uncondensed root, can in principle
lie anywhere on the portion of the real axis to the right
of the condensation pole at $z=1$. Their position, i.e. the value of
$w\in[1,\infty)$, obviously depends on the position of point $(x,y)$
on $\varGamma_\NW$. Thus the value of $w$ naturally parameterizes
the curve $\varGamma_\NW$ from the top contact point $(\kappa,0)$ to the
left one $(0,\kappa)$.

Summarizing, we arrive at the following alternative recipe for the derivation
of the arctic curve: on the basis of the two properties of the multiple
integral representation for EFP listed above,
we assume that they are determined  by the condition
of condensation of roots. Then, given the system of saddle-point equations
we impose condensation of `almost all' its solutions,
obtaining a  reduced saddle-point equation, determining the position of the
uncondensed roots. We then require the existence among them
of two coinciding roots with value $w\in[1,\infty)$,
which parameterizes the points on the arctic curve.

To illustrate and verify this recipe, we come back to saddle-point
equation, \eqref{SPEcondFF}, and implement the condensation of almost
all roots. As already explained, the last term of
LHS of \eqref{SPE_FF} in the large $s$ limit reduces to $2/(z_j-1)$.
As a result, this term combines with the first term and we
end with the  following reduced saddle-point equation
\begin{equation}\label{SPEcondFF}
\frac{y}{z-1}-\frac{1-x}{z}+\frac{(1-y)t^2}{1+ t^2 z}=0.
\end{equation}
This equation determines  the position of those roots which survive
in the complex plane and do not condense at $z=1$.

We now require  the existence among them
of two coinciding roots. Noting that the numerator of \eqref{SPEcondFF}
is of second order in $z$, we can  simply require the vanishing of
its discriminant, thus obtaining \eqref{arctic_ellipses}.
Equivalently, denoting  LHS of \eqref{SPEcondFF} by $F(z)$,
we can require it to be of the form  $F(z)=(z-w)^2\wt{F}(z)$, with
$\wt{F}(z)$ regular in the vicinity of $w$. This translates into
a system of two equations $F(w)=0$, $F'(w)=0$, linear in
its unknowns $x$ and $y$, with the solution
\begin{equation}\label{arctic_ellipses_param}
x=\frac{t^2 w^2}{t^2 w^2 +1},\qquad
y=\frac{t^2 (w^2 -2 w+1)}{(t^2+1)(t^2 w^2+1)},
\qquad w\in [1,\infty).
\end{equation}
It can be easily seen that, as $w$ varies over the interval $[1,\infty)$,
the point  $(x,y)$ above indeed describe the curve $\varGamma_\NW$,
between the two  contact points $(\kappa,0)$ and $(0,\kappa)$.
Elimination of parameter $w$
from expressions \eqref{arctic_ellipses_param} leads again  to
equation \eqref{arctic_ellipses} for the arctic ellipse.

\section{Arctic curves in the disordered regime}

\subsection{Condensation hypothesis}

In the previous Section we have shown that the assumption of
the correspondence  between condensation of roots and the arctic curve
allows one to derive this curve in a parametric form,
through the requirement of a pair of coinciding roots for the
reduced saddle-point equation. For the case considered there, the
correspondence   between condensation of roots and arctic curve
can be verified by direct inspection of the form of the Green function,
which indeed reduces to expression \eqref{Greeneq2} for
$(x,y)\in\varGamma_\NW$.

It turns out that in the general case, when the value of
$\Delta$ is arbitrary, the method explained in
Section 5.3 remains applicable! This observation is based on the
fact that the two properties of the
integrand of multiple integral representation for EFP
listed in Section 5.3 appear to be
totally independent of the value of $\Delta$.
This means that we can again assume that the arctic curve arises in
correspondence to condensation of almost all roots of the
saddle-point equations.
It is to be emphasized, that in the general case however
such correspondence cannot be verified, not even a posteriori
(at least by the methods at our disposal), hence
we call it here `condensation hypothesis'.

Running a few steps forward, it worth to mention here
that the only limitation which will restrict  the final result
to be valid only in  the disordered regime ($|\Delta|<1$) comes from
formula \eqref{hNlargeN} which will play an important role below
in the analysis of the reduced saddle-point equation.
For this reason we somehow restrict ourselves here to the disordered regime
although the discussion is in fact rather general.

As just said, the  `condensation hypothesis' is strongly supported by the
form of multiple integral representation \eqref{MIR2} for EFP in the
general case, satisfying the two crucial properties listed in Section
5.3. To show that the first property is indeed fulfilled, we have to
consider, for $r,s=1,\dots,N$, the integral
\begin{multline}\label{INrs2}
I_N^{(r,s)}=\frac{(-1)^{s(s+1)/2}Z_s}{s!(2\pi\rmi)^s a^{s(s-1)}c^s}
\oint_{C_1^-}^{} \cdots \oint_{C_1^-}^{}
\prod_{j=1}^{s} \frac{[(t^2-2\Delta t)z_j+1]^{s-1}}{z_j^r(z_j-1)^s}\,
\\ \times
\prod_{\substack{j,k=1\\ j\ne k}}^{s} \frac{1}{t^2 z_jz_k-2\Delta t z_j +1}
\prod_{1\leq j<k\leq s}^{} (z_k-z_j)^2
\\ \times
h_{N,s}(z_1,\dots,z_s)h_{s,s}(u_1,\dots,u_s)
\,\rmd z_1\cdots \rmd z_s.
\end{multline}
This integral differs from \eqref{MIR2} only in the integration contour
(recall that $C_1^-$ denotes the closed negative-oriented contour in
the complex plane, enclosing point $z=1$, and no other singularity of the
integrand). To show that quantity \eqref{INrs2} is identically equal
to one, let us resort to another, essentially equivalent, representation
\begin{multline}\label{INrs1}
I_N^{(r,s)} :
= \frac{(-1)^s}{(2\pi \rmi)^s}
\oint_{C_1^-}^{} \cdots \oint_{C_1^-}^{}
\prod_{j=1}^{s}\frac{[(t^2-2\Delta t)z_j+1]^{s-j}}{z_j^r(z_j-1)^{s-j+1}}\,
\\ \times
\prod_{1\leq j<k \leq s}^{} \frac{z_j-z_k}{t^2z_jz_k-2\Delta t z_j+1}\;
h_{N,s}(z_1,\dots,z_s)
\,\rmd z_1\cdots \rmd z_s.
\end{multline}
It is to be emphasized that equivalence of representations \eqref{INrs2}
and \eqref{INrs1} simply follows from the fact that the integrand of
\eqref{MIR2} is the symmetrized version of the integrand of \eqref{MIR1}
(see \cite{CP-07b} for details,
and also Appendix C of  \cite{KMST-02} where a key identity
has been proven). Performing integration in representation
\eqref{INrs1} in the variable $z_s$, and recalling property \eqref{hnszto1},
the identity
\begin{equation}\label{I=1}
I_N^{(r,s)}=1,\qquad  (r,s=1,\dots,N)
\end{equation}
follows immediately by induction. In this way we observe that the first
property, which states that the cumulative residue of the integrand for EFP
over all variables $z_1,\dots,z_s$ at the pole $z=1$ is equal to $1$, is
indeed
fulfilled for arbitrary value of $\Delta$.

The second property, that the pole at $z=1$ for every integration
variable $z_1,\dots,z_s$ is of order $s$, follows by simple inspection of
representation \eqref{MIR2}. This means that we can again expect
condensation of almost all roots of the saddle-point equations at $z=1$.
Note that the other poles in the integrand of representation
\eqref{MIR2}, being poles of small order even in the scaling limit, can
only give subleading contributions, with respect to the stepwise
behaviour generated by condensation at $z=1$. It is to be emphasized that
although in the general situation we have no connection with the matrix
model formulation any more, we can nevertheless represent the integrand
in the exponential form with an `action' which will contain, among many
others terms, the term $s\sum_{j=1}^{s} \ln(z_j-1)$. Assuming validity of
the condensation hypothesis we thus assume that this term still dominates
when the parameters ($x$ and $y$) are near the arctic curve and hence
other terms have no relevance for the mechanism of condensation (although
these other terms can contribute to the reduced saddle-point equation and
therefore determine the arctic curve).

Thus we have just seen that the two main properties of the multiple
integral representation \eqref{MIR2} for EFP are fulfilled,
strongly supporting the condensation hypothesis.
Hence, we may apply the procedure explained in Section 5.3.
This will be done in the remaining part of this section. Namely,
we will derive the corresponding
reduced saddle-point equation, and implement the condition of a pair
of coinciding roots, to obtain the arctic curve in a parametric form.

\subsection{`Reduced' saddle-point equation}

We thus assume condensation of solutions of saddle-point equations
\eqref{SPE},
and derive the corresponding reduced saddle-point equation.
We start with setting $n_c$ of the $s$ variables $z_k$, $k=1,\dots,s$,
to the value $1$.  We are left with a system of $n_u$ equations in
the $n_u$ uncondensed  variables, let them be $z_j$, $j=1,\dots,n_u$. In what
follows, the
fact that $n_u/s$ vanishes in the scaling limit, see \eqref{n_u}, plays a
crucial
role.

In  saddle-point equation \eqref{SPE}, let us
consider the last term in the sum.
As explained in Section 5.3, condensation of almost all  roots reduces it
to $2s/(z_j-1)$, at leading order for large $s$. Similarly, for the remaining
two terms in this sum, for large $s$, we have, respectively,
\begin{equation}
\sum_{\substack{k=1\\ k\ne j}}^{s}
\frac{ t^2 z_k-2\Delta t}{t^2z_jz_k-2\Delta t z_j +1}
\longrightarrow
s \frac{(t^2-2\Delta t)}{(t^2-2\Delta t )z_j+1}+o(s),
\end{equation}
and
\begin{equation}
\sum_{\substack{k=1\\ k\ne j}}^{s}
\frac{t^2 z_k}{t^2z_jz_k-2\Delta t z_k +1}
\longrightarrow
s \frac{t^2}{t^2 z_j -2\Delta t +1}+o(s).
\end{equation}
Further, due to property
\eqref{hnszto1} function $h_{N,s}(z_1,\dots,z_s)$, appearing in the term
after the sum in equation \eqref{SPE}, simplifies to function
$h_{N,n_u}(z_1,\dots,z_{n_u})$, which, in turn, for $N,s\to\infty$, and
$n_u/N\sim 0$, can be evaluated for large $s$ directly from its definition
\eqref{hNs-def}. In this way we obtain
\begin{align}
\ln h_{N,s}(z_1,\dots,z_s)
&\longrightarrow
\ln h_{N,n_u}(z_1,\dots,z_{n_u})
\notag\\ & \quad\qquad
=\sum_{j=1}^{n_u} \ln h_N(z_j) +o(s).
\end{align}
Similarly, recalling that $u_k\to 0$ as $z_j\to 1$, see \eqref{uofz},
and using property \eqref{hnszto0}, we find
that function $h_{s,s}(u_1,\dots,u_s)$, appearing in the last term of
equation \eqref{SPE},
simplifies, modulo an unessential factor, to function
$h_{n_u,n_u}(u_1,\dots,u_{n_u})$.
Recalling that
$h_{n_u,n_u}(u_1,\dots,u_{n_u})$ is a polynomial of order $n_u$ in each
of its variables, we obtain that its logarithm for large $s$
is estimated as $o(s)$. Thus we have
\begin{align}
\ln h_{s,s}(u_1,\dots,u_{s}) &\longrightarrow
\ln h_{n_u,n_u}(u_1,\dots,u_{n_u})+
\sum_{j=n_u+1}^s \ln h_j(0)
\notag\\ &\quad\qquad
= C_1 s^2  + C_2 s +o(s).
\end{align}
Here  $C_1$ and $C_2$ are some quantities which do not
depend on $u_j$ $(j=1,\dots,n_u)$. After differentiating,
we are left with a term estimated as $o(s)$, which is to be neglected, at the
leading order in the large $s$ limit.

As a result, we obtain that saddle-point equations \eqref{SPE} simplify to
a  set of $n_u$ identical and decoupled equations,
\begin{equation}\label{RSPE}
F(z_j) =0,\qquad (j=1,\dots,n_u).
\end{equation}
Here function $F(z)$ is given by
\begin{equation}\label{Flarge}
F(z):=
\frac{y}{z-1}-\frac{1-x}{z}-\frac{y t^2}{t^2 z-2\Delta t +1}
+\left(\lim_{N\to\infty}\frac{\ln h_N(z)}{N}\right)',
\end{equation}
where  the scaling variables $x$ and $y$ are defined in \eqref{scalingvar}.

To write the explicit form of the equation determining
the location of uncondensed saddle-point solutions, we need to know the last
term
in the expression for $F(z)$.
In what follows we shall need
to consider the case of $z$ real and positive. For these values we can use
formula \eqref{hNlargeN} for large
$N$ limit of function $h_N(z)$. From this formula, for the last term in
\eqref{Flarge}
we obtain
\begin{equation}
\lim_{N\to\infty} \frac{\ln h_N(\gamma(\xi))}{N}
=\ln\left(\frac{\sin\alpha(\lambda-\eta)\sin\alpha\xi\sin(\xi+\lambda-\eta)}
{\alpha\sin(\lambda-\eta)\sin\xi\sin\alpha(\xi+\lambda-\eta)}
\right).
\end{equation}
Taking into account that function $\gamma(\xi)$, see \eqref{gamma},
is a monotonously growing function, from $-\infty$ to $+\infty$, over the
interval
$\xi\in (-\lambda-\eta,\pi-\lambda-\eta)$, which is
the interval of periodicity of $\gamma(\xi)$ (note that in fact
$\gamma(\xi)$ is a rational function of $\cot\xi$),
we can switch from function $F(z)$ to function $f(\xi)$, defined by
\begin{equation}\label{Fgammaxifxi}
F(\gamma(\xi))=:
\frac{\sin(\lambda-\eta)\sin\!^2(\xi+\lambda+\eta)}{\sin(\lambda+\eta)\sin
2\eta}\, f(\xi).
\end{equation}
Here the prefactor in fact is equal to $1/\gamma'(\xi)$.
Direct calculation from \eqref{RSPE} lead us to the following neat
expression for this function
\begin{equation}\label{fxi}
f(\xi)= x\,\varphi(\xi+\lambda) + y\,\varphi(\xi+\eta) - \varPsi (\xi),
\end{equation}
where function $\varphi(\xi)$ is given by \eqref{varphi} and function
$\varPsi(\xi)$ reads
\begin{equation}
\varPsi(\xi):=\cot\xi-\cot(\xi+\lambda+\eta)-\alpha\cot(\alpha\xi)+
\alpha\cot\alpha(\xi+\lambda-\eta).
\end{equation}
Noting properties $\varPsi(\pi-\lambda-\eta-\xi)=\varPsi(\xi)$ and
$\varphi(\pi-\xi)=\varphi(\xi)$,
it easy to see that function $f(\xi)$ obeys the symmetry under the
substitution
$\xi\mapsto \pi-\lambda-\eta-\xi$ and
simultaneous interchange of the coordinates, $x\leftrightarrow y$. This
property
of function $f(\xi)$ will lead to the diagonal-reflection symmetry of the
arctic curve.

\subsection{The arctic curve}

To obtain the arctic  curve from the reduced saddle-point equation, we
will follow the recipe discussed in detail in Section 5.3. We thus
have to require that function $F(z)$ has a double zero, which, moreover,
must lie on the interval $[1,\infty)$. Denoting the value of this zero by
$w$, we therefore require that $F(z)=(z-w)^2 \tilde F(z)$, with
$\tilde F(z)$ regular in the vicinity of $w$.
This condition is equivalent to the system of two equations
\begin{equation}\label{FF'}
F(w)=0, \qquad F'(w)=0.
\end{equation}
These are equations for unknown $x$ and $y$, which are functions of $w$.
For each value of $w\in[1,\infty)$, values of $x$ and $y$ correspond to a
point of the arctic curve, i.e., the solution of system \eqref{FF'}  is
just the arctic curve in a parametric form, with $w$ being the parameter.

Taking into account the observation above that instead of function
$F(z)$ we can use function $f(\xi)$, introduced by formula
\eqref{Fgammaxifxi}, we immediately obtain that
system of equations \eqref{FF'} is equivalent to
the following system of equations
\begin{equation}
f(\zeta)=0,\qquad f'(\zeta)=0,
\end{equation}
where $\zeta$ is the new parameter which parameterizes the arctic curve,
$w=\gamma(\zeta)$. The range of values of the original parameter
$w$, varying in the interval $[1,\infty)$, corresponds
$\zeta\in[0,\pi-\lambda-\eta]$.

We now solve the linear system of
the two equations above, and arrive at the
following result:
\begin{equation}\label{arctic}
\begin{split}
x=\frac{\varphi'(\zeta+\eta)\varPsi(\zeta)-\varphi(\zeta+\eta)\varPsi'(\zeta)}
{\varphi(\zeta+\lambda)\varphi'(\zeta+\eta)-\varphi(\zeta+\eta)
\varphi'(\zeta+\lambda)},
\\
y=\frac{\varphi(\zeta+\lambda)\varPsi'(\zeta)-\varphi'(\zeta+\lambda)
\varPsi(\zeta)}
{\varphi(\zeta+\lambda)\varphi'(\zeta+\eta)-\varphi(\zeta+\eta)
\varphi'(\zeta+\lambda)}.
\end{split}
\end{equation}
These two equations constitute
the parametric form of the top left portion, $\varGamma_\NW$,
of the arctic curve, as the parameter
$\zeta$ varies in the interval $[0,\pi-\lambda-\eta]$. The value
$\zeta=0$ corresponds to the contact point of
the curve $\varGamma_\NW$ with the $x$-axis, and as $\zeta$ increases,
the whole curve $\varGamma_\NW$ is monotonously constructed, up to
the contact point with the $y$-axis, at $\zeta=\pi-\lambda-\eta$.

Using the properties of the functions involved here, we can write the result
in the form
\begin{equation}\label{XYzeta}
x=X(\zeta),\qquad y=Y(\zeta),\qquad
\zeta\in[0,\pi-\lambda-\eta],
\end{equation}
where functions $X(\zeta)=X(\zeta;\lambda,\eta)$ and
$Y(\zeta)=Y(\zeta;\lambda,\eta)$
are simply related by
\begin{equation}\label{XandY}
X(\zeta)=Y(\pi-\lambda-\eta-\zeta).
\end{equation}
Direct calculation gives
\begin{align}\label{Yzeta}
Y(\zeta)&
=\frac{\sin\!^2\zeta\sin\!^2(\zeta+2\eta)\sin(\zeta+\lambda-\eta)
\sin(\zeta+\lambda+\eta)}
{\sin2\eta\sin(\lambda-\eta)
\big[\sin(\zeta+\lambda-\eta)\sin\zeta+\sin(\zeta+\lambda+\eta)\sin(\zeta+2\eta)\big]}
\notag\\ &\quad\times
\bigg\{
\frac{\sin(\lambda-\eta)\sin(\lambda+\eta)}
{\sin\!^2\zeta\sin(\zeta+\lambda+\eta)\sin(\zeta+\lambda-\eta)}
\notag\\ &\qquad\quad
+\frac{\sin(2\zeta+2\lambda)}{\sin(\zeta+\lambda-\eta)\sin(\zeta+\lambda+\eta)}\,
\frac{\alpha\sin\alpha(\lambda-\eta)}{\sin\alpha\zeta\sin\alpha(\zeta+\lambda-\eta)}
\notag\\ &\qquad\quad
-\frac{\alpha^2\sin\alpha(2\zeta+\lambda-\eta)\sin\alpha(\lambda-\eta)}
{\sin\!^2\alpha\zeta\sin\!^2\alpha(\zeta+\lambda-\eta)}
\bigg\}.
\end{align}
Formulae \eqref{XYzeta}--\eqref{Yzeta}
represent the main result of the present paper.

As a comment to the result, it is worth to mention that in all examples discussed in
the literature to date, the arctic curves (or frozen boundaries of limit
shapes) always turn out to be algebraic curves. From the form of function
$Y(\zeta)$ above it is straightforward to conclude that as far as the
parameter $\alpha=\pi/(\pi-2\eta)$ is irrational, the arctic curve of the
domain-wall six-vertex model is a non-algebraic curve. This is the
situation for generic weights of the six-vertex model in the disordered
regime. On the other hand, the arctic curve is  algebraic when $\alpha$ is a
rational number, since then both $\cot\zeta$ and $\cot\alpha\zeta$,
rationally entering in \eqref{Yzeta}, can be expressed as rational
functions of a suitably chosen parameter.
In other words, the arctic curve in the domain-wall six-vertex model is
rational only when the weights correspond to the so-called root-of-unity
cases.

\section{Particular cases and combinatorial applications}

\subsection{Alternating sign matrices}

The arctic curve, whose general expression has been
given in the previous Section, for arbitrary weights
of the six-vertex model with domain wall boundary conditions in its
disordered regime, is worth to be investigated in more detail
when some of the parameters of the model are specialized to certain values.
In this section we consider a few  particular cases, some of which have a natural
interpretation in the context of the limit shape of large
alternating sign matrices. Indeed, our result for the arctic curve of the
six-vertex model allows us to obtain the limit shapes of
alternating sign matrices within $q$-enumeration schemes (for the
disordered regime $0<q\leq 4$, see below).

We recall that an alternating sign matrix is a matrix which has only
$1$'s, $0$'s and $-1$'s in its entries, obeying, moreover, the rule
that in each row and each column of the matrix all
nonzero entries alternate in sign, and the first and the last nonzero
entries are $1$'s. In $q$-enumeration, denoted here as $A_N(q)$,
each matrix has weight $q^k$ where
$k$ is the number of $-1$'s in its entries.
For certain values of $q$, namely $q=1,2$, and $3$, the numbers of
$q$-enumerated alternating sign matrices,
$A_N(q)$, are known to be given by some factorized explicit expressions.
These results, stated first as conjectures, remained challenging for their
proofs for a long time, see  \cite{EKLP-92,Ze-96,Ku-96} and references
therein; the story and backgrounds of the problem
can also be found in book \cite{Br-99}.

A possible approach to alternating sign matrices exploits
their close relationship with the domain-wall six-vertex model.
Indeed, in \cite{EKLP-92} it was noticed that there is
a one-to-one correspondence between $N$-by-$N$ alternating sign matrices
and configurations of the domain-wall six-vertex model on the $N$-by-$N$
lattice.
Due to this correspondence,  $A_N(q)$ is equal, modulo a simple factor,
to $Z_N$, the partition function
of the domain-wall six-vertex model. The weights of this model have to
satisfy the relations $a=b$ and, since $c$-weights comes in pairs, $c=a\sqrt q$.
The first relation is fulfilled when
$\lambda=\pi/2$ in \eqref{param}, and the second one implies that
$q$ and $\Delta$ (recall that $\Delta=\cos2\eta$) are related by
\begin{equation}\label{q-eta}
q=2-2\Delta.
\end{equation}
The precise relation between
$q$-enumeration of $N$-by-$N$ alternating sign matrices,
$A_{N}(q)$, and the partition function $Z_N$ (at $\lambda=\pi/2$) is
\begin{equation}\label{ZNAN}
Z_N=a^{N^2} q^{N/2} A_{N}(q),
\end{equation}
see, e.g., \cite{EKLP-92} for a discussion.

In the context of alternating sign matrices, the arctic curve of the domain-wall
six-vertex model describes what is usually referred to as the `limit shape'.
Indeed, in their corner regions, alternating sign matrices mostly contain
$0$'s,  while in the interior they have  many nonzero entries;
as the size of matrices increases,
the probability of finding $1$'s and $-1$'s in the corner regions
vanishes, while in the central region such probability remains finite \cite{P-01}.
As a result, in the scaling limit (i.e., when large matrices
are scaled to a unit square) the arctic curve of the domain-wall six-vertex model
precisely describes the shape of this central region
(which is nothing but the region $\DI$, see Section 4).

A derivation of the limit shape of large $q$-enumerated alternating sign
matrices at $q=1$ and $q=3$ (in addition to the well-known case of $q=2$)
was given in our previous paper \cite{CP-08}. There we used the
observation of \cite{CP-05b} that at $\eta=\pi/6,\pi/4,\pi/3$ (with
$\lambda=\pi/2$), corresponding to the cases of $q=1,2,3$,
respectively, the H\"ankel determinant standing in formula \eqref{DN} for $Z_N$
turns out to be related with certain classical orthogonal
polynomials (this explains, additionally, the `roundness' of $A_N(q)$ at
$q=1,2,3$). At these values of parameters the function $h_N(z)$ turns out to
be expressible in terms of hypergeometric series and
its large $N$ asymptotic behaviour can be found directly, thus allowing us
to derive the arctic curve on the basis of the `condensation hypothesis'.
Below we discuss several particular cases, and, in particular,
how the results given in \cite{CP-08} follow from formulae
\eqref{XYzeta}--\eqref{Yzeta}, describing the arctic curve.

\subsection{Particular cases}

Let us consider the arctic curve for some particular values of
weights of the disordered regime. As mentioned above, if the
parameter $\lambda$ is specialized to $\pi/2$, then the
arctic curve also describes the limit shape of large
alternating sign matrices, within the corresponding $q$-enumeration scheme. We
consider here the cases which correspond to $q=1,2,3$, and $q=4$ where this
last case is obtained as a limiting case, as $\Delta\to -1$. We also
consider another limiting case, where $q$ vanishes. Interestingly, in
this case the limit shape tends to certain nontrivial limiting curve, rather than
becoming somehow degenerate or trivial; such limiting curve is
discussed in more details below.

\subsubsection{The case $\Delta=0$}

We start with the case  $\Delta=0$, or $\eta=\pi/4$. In this case, of
course, we expect to reproduce the arctic ellipse discussed in detail in
Section 5. Indeed, setting $\eta=\pi/4$ in formula \eqref{Yzeta}, we
obtain
\begin{equation}\label{ffcase}
Y(\zeta)=\frac{1}{2}(1-\cos2\zeta), \qquad \zeta \in \big[0,
\tfrac{3\pi}{4}-\lambda\big].
\end{equation}
We also have $X(\zeta)=\tfrac{1}{2}(1+\sin(2\zeta+2\lambda))$, and we can easily
eliminate parameter $\zeta$ in equations \eqref{XYzeta}. As a result,
we arrive again to equation \eqref{arctic_ellipses}, where
$t:=\tan(\lambda-\pi/4)$.

\subsubsection{The case $\Delta=\frac{1}{2}$}

The case $\Delta=1/2$, or $\eta=\pi/6$, is interesting since at
$\lambda=\pi/2$ the model is equivalent to the enumeration of
alternating sign matrices (with $q=1$). Specifying $\eta$
to the value $\pi/6$ and setting $\lambda=\pi/2$, we find
that expression \eqref{Yzeta} simplifies to
\begin{equation}
Y(\zeta)=1-\cos\zeta,\qquad \zeta\in \big[0,\tfrac{\pi}{3}\big].
\end{equation}
Correspondingly, we also have $X(\zeta)=1-\cos(\pi/3-\zeta)$ and, eliminating
parameter $\zeta$, it can be found that the curve $\varGamma_\NW$ (the top-left
portion of the arctic curve) is described by the equation
\begin{equation}\label{icecurve}
(2x-1)^2+(2y-1)^2-4xy=1\,, \qquad x,y\in\big[0,\tfrac{1}{2}\big].
\end{equation}
This curve describes the limit shape of large alternating signs matrices
\cite{CP-08}. Interestingly, in comparison with the arctic circle, given by equation
$(2y-1)^2+(2x-1)^2=1$, it has just single additional term $-4xy$ in LHS.
The property of the curve $\varGamma_\NW$ to be given by a quadratic equation in
the $\eta=\pi/6$ case holds only at $\lambda=\pi/2$. Indeed,
just specifying $\eta=\pi/6$ but keeping $\lambda$
generic, one finds from \eqref{Yzeta} that function $Y(\zeta)$
in this case is rather bulky;
the curve $\varGamma_\NW$ turns out to be described by a tenth order equation.

\subsubsection{The case $\Delta=-\frac{1}{2}$}

The case of $\Delta=-1/2$, or $\eta=\pi/3$, will be treated here only at
$\lambda=\pi/2$. At this choice of parameters the model
is equivalent to $q$-enumeration of alternating sign
matrices with $q=3$, which is a well-known example of tractable
enumeration (together with the cases $q=1,2$). Specifying in \eqref{Yzeta}
$\eta=\pi/3$ and $\lambda=\pi/2$, we obtain
\begin{equation}\label{dualcurve}
Y(\zeta)=4
\left[\frac{\sin(\tfrac{\pi}{3}-\zeta)\tan\zeta}{1+2\cos2\zeta}\right]^2
\frac{11+12\cos2\zeta-2\cos4\zeta}{6-3\cos2\zeta-\sqrt3\sin2\zeta},
\qquad \zeta\in\big[0,\tfrac{\pi}{6}\big].
\end{equation}
We also have $X(\zeta)=Y(\pi/6-\zeta)$. For further analysis,
it is convenient to rewrite  the parametric formulae for the curve
in terms of rational functions of a suitably chosen parameter.
For example, choosing the parameter $w=\sin(\pi/6+\zeta)/\sin(\pi/6-\zeta)$,
one obtains formulae for the curve given in \cite{CP-08}.
Further, excluding the parameter $w$ one can find the corresponding
algebraic equation for the curve, which appears to be
of the sixth order (see \cite{CP-08}, equation (14)).

\subsubsection{The case $\Delta= -1$}

This case can be obtained as the limit from the
disordered regime; in fact, the case of $\Delta= -1$ is often regarded as
belonging to the disordered regime since the model is still
disordered (at $\Delta=-1$ the model undergoes an infinite order phase
transition). Denoting by $v$ the rapidity variable of the model at
$\Delta=-1$, the standard parameterization of the weights of this model
reads
\begin{equation}
a=1+v,\qquad b=1-v,\qquad c=2 \qquad (-1<v<1).
\end{equation}
To fit this parametrization, one can perform the scaling in the parameters
by taking the limit $\delta\to 0$, upon setting (see equation \eqref{param})
\begin{equation}\label{delta}
\lambda=\frac{\pi}{2}-v\delta,\qquad \eta=\frac{\pi}{2}-\delta.
\end{equation}
Correspondingly, we also set $\zeta=p\delta$ and then equations
\eqref{XYzeta} for the arctic curve
will read $x=\wt X(p)$ and $y=\wt Y(p)$, where
$p$ is the new parameter which parameterizes the curve,
$p\in[0,1+v]$, and
\begin{equation}
\wt X(p)=\lim_{\delta\to0}X(p\delta),\qquad
\wt Y(p)=\lim_{\delta\to0}Y(p\delta),
\end{equation}
where $\delta$ also enters $\lambda$ and $\eta$ as given by \eqref{delta}.
Functions $\wt X(p)$ and $\wt Y(p)$ also depend on $v$ as a parameter
and, due to \eqref{XandY}, they satisfy $\wt X(p)= \wt Y(1+v-p)$.
Explicitly, function $\wt Y(p)$ reads
\begin{multline}
\wt Y(p)=\frac{(2-p)^2}{4(1-v)\big[(1+v)(1-p)+p^2\big]}
\Bigg\{1-v^2
- \pi p^2(p-v)\,\frac{\cos\frac{\pi}{2}v}
{\sin\frac{\pi}{2} p\,\cos\frac{\pi}{2}(p-v)}
\\
-\pi^2 p^2\big[1-(p-v)^2\big]\frac{\cos\frac{\pi}{2} v \; \cos\frac{\pi}{2}(2p-v)}
{4\sin\!^2\frac{\pi}{2}p\,\cos\!^2\frac{\pi}{2}(p-v)}\Bigg\},
\end{multline}
where $p\in[0,1+v]$. Clearly, in this case, contrarily to the three
examples considered above, the arctic curve is not an algebraic one.
In the case of $v=0$, that is, when the weights
satisfy $a=b$, the arctic curve describes the limit
shape of large alternating sign matrices within the $q$-enumeration at $q=4$.

\subsubsection{The case $\Delta\to 1$}

This case corresponds to $\eta\to 0$. For arbitrary $\lambda$ and
small $\eta$ formula \eqref{Yzeta} reads
\begin{equation}
Y(\zeta)=\frac{1}{2\pi}(2\zeta-\sin2\zeta)+O(\eta),\qquad
\zeta\in[0,\pi-\lambda-\eta].
\end{equation}
Correspondingly, we also have $X(\zeta)=Y(\pi-\lambda-\eta)$.
The subsequent  analysis depends on whether or not, as $\eta$ vanishes,
the parameter $\lambda$ is scaled accordingly.

Namely, the first possibility is to
set $\lambda=\eta v$ or $\lambda=\pi-\eta v$
where $v$ is a new rapidity variable ($v\geq 1$); this choice corresponds
to approaching either of the two branches, $a>b$ or $a<b$, respectively,
of the model at $\Delta=1$. The weights of the model at $\Delta=1$ are
parameterized as $a=v\pm1$, $b=v\mp1$, and $c=2$, where $v\geq 1$.
The two choices of the signs corresponds to the two branches of the
model at $\Delta=1$. At $a>b$, we find, after eliminating the parameter
$\zeta$, that the curve $\varGamma_\NW$ is just the straight line:
$x+y=1$, where $x,y\in[0,1]$. At $a<b$ the curve $\varGamma_\NW$ is just
single point: $x=y=0$. All this is in agreement with the fact that at
$\Delta=1$ the model is not disordered anymore; the region $\DI$, see
figure \ref{fig-arctic}, degenerates into the straight line $x+y=1$ (if
$a>b$) or $x=y$ (if $a<b$).

The second possibility, which appears to be also the most interesting,
is to keep $\lambda$ fixed as $\eta$ vanishes.
In the phase diagram of the six-vertex model in the $a/c$--$b/c$ plane
this corresponds to taking the limit into the deep infinity of the disordered
region, rather than approaching either of the two branches
of the $\Delta=1$ model. In particular, setting $\lambda=\pi/2$ and neglecting
small $\eta$ corrections, one can easily find that the curve
$\varGamma_\NW$ is given by the equation
\begin{equation}\label{zeroenum}
x+y=\frac{1}{2}-\frac{1}{\pi}\cos\pi(x-y),\qquad
x,y\in\big[0,\tfrac{1}{2}\big].
\end{equation}
This equation has an interesting meaning in the context of alternating sign
matrices: it is the limiting curve describing the limit shape as $q$ tends to zero.

\subsection{Discussion}

A natural question concerns the qualitative behaviour of arctic curve
\eqref{XYzeta}--\eqref{Yzeta} as one varies parameters of the model.
In discussing some properties of the arctic curve it is useful to resort
to numerical plotting. For example,
considering the case of generic $\lambda$, we just mention here that as
$\lambda$ varies,
the curve is deformed along one of the two diagonals
according to the sign of $\lambda-\pi/2$. An example of the whole
arctic curve $\arctic$ in a non-symmetric case is shown in
figure \ref{fig-arctic}. Namely, this figure shows the arctic curve
at $\eta=\pi/6$ and $\lambda=5\pi/12$.

\begin{figure}
\centering
\input{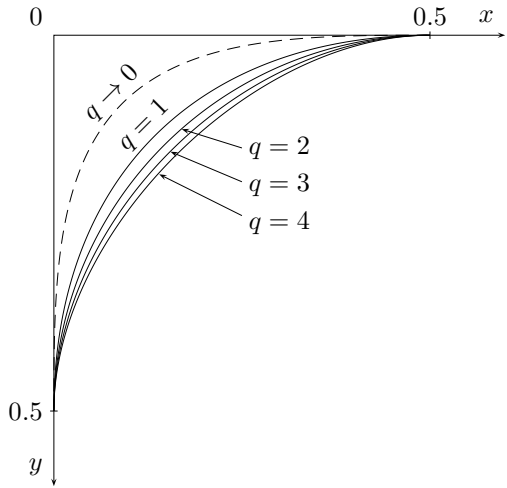}
\caption{The limit shapes of the large alternating sign matrices within
various $q$-enumeration schemes. As $q$ vanishes, the limit shape tends
to a limiting  curve (dashed line), given by equation \protect\eqref{zeroenum}.}
\label{fig-shapes}
\end{figure}

Specializing to the case of $\lambda=\pi/2$ we can also discuss
the arctic curve in application to the limit shape of large $q$-enumerated
alternating sign matrices, with $q$ and $\eta$ related by \eqref{q-eta}.
The relevance  of formulae \eqref{XYzeta}--\eqref{Yzeta} in
the context of alternating sign matrices is that they allow one to study the
variation of the limit shape as $q$ varies over the interval $(0,4]$. In
addition to the cases of $q=1,2,3$ considered in \cite{CP-08}, in
figure \ref{fig-shapes} we plot also the limiting cases of $q=4$ and $q\to 0$
(the latter shown by a dashed line).  Note that, as expected, the disordered region
(the area enclosed by the limit shape) is largest for
$q\to 0$, and slowly decreases as $q$ increases over the considered interval.

Comparing numerical plots of the arctic curve at various values of
the parameters of the model we find that our results are
completely compatible with all available numerical data
\cite{SZ-04,AR-05},  which are however affected by large uncertainties, due to
the huge  technical difficulties in  this kind of computer simulations.
The most refined computer simulations available at the moment
has been performed by Ben Wieland for  the especially relevant case
of alternating sign matrices at $q=1$, corresponding to the value $\Delta=1/2$.
Pictures comparing these numerical data with the corresponding arctic curve,
given by equation \eqref{icecurve}, are available \cite{W}.

Coming back to the qualitative behaviour of the arctic curve, we would like
to focus again on  the limiting  curve, which, in the context of limit
shapes of large alternating sign matrices, is referred above as the
`$q\to 0$' curve. Note that such a curve exists for any fixed value of
$\lambda$, as $\eta\to 0$ (see discussion in Section 7.2.5).
Concerning  the emergence of a non-trivial limiting curve, in hindsight,
it is clear that this is ascribable to the fact that the two limits,
$N\to \infty$, and $q\to 0$, do not commute.

\section{Conclusions}

In the present paper we have derived the arctic curve for the six-vertex
model with domain wall boundary conditions, in its disordered regime. The
derivation is essentially based on the exact expression in terms of a
multiple integral for the emptiness formation probability, a
correlation function which discriminates order and disorder. We have
observed that in the scaling limit the arctic curve can be obtained as
the  condition that almost all solutions of the system of coupled
saddle-point equations for the multiple integral representation of
emptiness formation  probability condense at the same, known, value.
The explicit expression for the curve in parametric form
shows that in general it is a non-algebraic curve; for special choices of
weights of the model (the so-called root-of-unity cases)
it simplifies to algebraic curves. We have also discussed
combinatorial applications of the result to the problem of limit shapes
of large alternating sign matrices within various enumeration schemes. In
particular, we find that the limit shape of $q$-enumerated
alternating sign matrices has a non-trivial limit as $q$ tends to zero;
furthermore, such limiting curve is described by a very simple
equation.

Having established the expression of the arctic curve, a natural question
concerns its fluctuations, which, in our approach, are related to the
subleading corrections to the stepwise behaviour of emptiness formation
probability in the scaling
limit. In the case of domino tiling of Aztec diamond the fluctuations of
the arctic circle are  governed by the Tracy-Widom distribution and the
Airy process \cite{J-02,J-05}. This results, which naturally extends to
the case $\Delta=0$ of the domain-wall six-vertex model \cite{FS-06},
appears rather natural in view of the `random matrix model' derivation of
the arctic circle, provided in \cite{CP-07a}. Indeed, the fluctuations of
the arctic curve are expected to be governed by the fluctuations of the
first eigenvalues evaporating from the logarithmic well where total
condensation occurs, and these  fluctuations are in turn  known from the
literature on Penner models (see, e.g., \cite{AMK-94}) to be governed by
the Tracy-Widom distribution. From the discussion of Sections 5 and 6,
this pictures extends rather naturally to generic values of $\Delta$ in
the disordered regime. On the basis of universality considerations, it is
thus very tempting to argue that fluctuations of the arctic curve are
still  governed by the Airy process, at least in the disordered regime. A
direct proof of this statement would be of great interest.

Another natural question concerns the extension of our results to the
anti-ferroelectric regime, $\Delta<-1$. We recall that in this case (see
Section 2.3) there are two different phase-separation curves, an outer
one, which is the usual arctic curve, separating regions of ferroelectric
order from an intermediate region of disorder, and an inner one,
separating this region of disorder from a central region of
anti-ferroelectric order. Since the emptiness formation probability
detects spatial transition from order to disorder,  the present approach
can be used to address the problem of the outer phase-separation curve.
Most of its ingredients  are independent of the value of $\Delta$, and
thus it can be directly  applied to the anti-ferroelectric regime. In
particular, in the special case of $\Delta\to-\infty$ (which is
technically similar to the free-fermion case) the arctic curve can be
readily derived, reproducing the result  of paper \cite{Zj-02}. For
generic values of $\Delta$ in the anti-ferroelectric regime, the only
open problem concerns the evaluation of the thermodynamic limit of
function $h_N(z)$. Such evaluation would provide the solution to the
problem of the outer phase-separation curve of the domain-wall six-vertex
model for the whole anti-ferroelectric regime.

\section*{Acknowledgments}

We thank Galileo Galilei Institute for Theoretical Physics for kind
hospitality during the completion of this work.
F.C. acknowledges partial
support from MIUR, PRIN grant 2007JHLPEZ, and from the European Science
Foundation program INSTANS.
A.G.P. acknowledges partial support from
INFN, Sezione di Firenze, from the Russian Foundation for Basic Research,
grant 07-01-00358, and from the programme ``Mathematical Methods in
Nonlinear Dynamics'' of Russian Academy of Sciences.

\appendix
\section{Partition function of partially inhomogeneous model and functions
$h_{N,s}(z_1,\dots,z_s)$}

Formula \eqref{ZN-Hankel} is a special case of a more general
determinant representation, known as Izergin-Korepin formula,
which was originally derived for the model with inhomogeneous weights
\cite{I-87}.
The weights are made position-dependent by attaching rapidity variables
to each vertical and horizontal line, so that there are $2N$
rapidity variables in total instead of just one variable $\lambda$.
Namely, the weights of the vertex lying at intersection of $i$th vertical
and $k$th horizontal lines (we enumerate lines
from right to left and from top to bottom) are parameterized as
\begin{equation}
a_{ik}=\sin(\lambda_i-\nu_k+\eta),\qquad
b_{ik}=\sin(\lambda_i-\nu_k-\eta),\qquad
c_{ik}=\sin2\eta.
\end{equation}
Correspondingly, the partition function is now a function of $2N$ rapidity
variables
$\lambda_1,\dots,\lambda_N,\nu_1,\dots,\nu_N$. In \cite{I-87} the
following representation was shown to be valid
\begin{multline}\label{IKformula}
Z_N(\lambda_1,\dots,\lambda_N;\nu_1,\dots,\nu_N)=
\frac{\prod_{i,k=1}^N \sin(\lambda_i-\nu_k+\eta)\sin(\lambda_i-\nu_k-\eta)}
{\prod_{1\leq i<k\leq N}
\sin(\lambda_i-\lambda_k)
\sin(\nu_k-\nu_i)}\\
\times\det M,
\end{multline}
where $M$ is an $N$-by-$N$ matrix with the entries
\begin{equation}
M_{ik}:=\varphi(\lambda_i-\nu_k)=\frac{\sin2\eta}{\sin(\lambda_i-\nu_k+\eta)
\sin(\lambda_i-\nu_k-\eta)}.
\end{equation}
To obtain \eqref{ZN-Hankel} from \eqref{IKformula}, one has to set
$\nu_k=0$ and $\lambda_i=\lambda$ ($i,k=1,\dots,N$). Due to the singularities
in the denominator of representation \eqref{IKformula}, this has to be
implemented as a limit, to be evaluated using l'H\^opital's rule.

In performing the limiting procedure to the
homogeneous model there are some interesting intermediate situations.
An important example is when the limit is performed only in
one set of variables, say, $\nu_k\to 0$ ($k=1,\dots,N$) while
all $\lambda_i$'s are kept generic (and different from each other).
Therefore, the weights are given by
\begin{equation}
a_{i}=\sin(\lambda_i+\eta),\qquad
b_{i}=\sin(\lambda_i-\eta),\qquad
c_{i}=\sin2\eta.
\end{equation}
and the partition function
$Z_N(\lambda_1,\dots,\lambda_N):=Z_N(\lambda_1,\dots,\lambda_N;0,\dots,0)$
reads
\begin{multline}\label{ZNpartial}
Z_N(\lambda_1,\dots,\lambda_N)=
\frac{\prod_{i=1}^{N}[\sin(\lambda_i+\eta)\sin(\lambda_i-\eta)]^N}
{\left(\prod_{n=1}^{N-1}n!\right)\prod_{1\leq i<k \leq
N}^{}\sin(\lambda_k-\lambda_i)}
\\ \times
\begin{vmatrix}
\varphi(\lambda_1)& \dots & \varphi(\lambda_N)\\
\varphi'(\lambda_1)& \dots & \varphi'(\lambda_N)\\
\vdots & \ddots & \vdots \\
\varphi^{(N-1)}(\lambda_1)& \dots & \varphi^{(N-1)}(\lambda_N)
\end{vmatrix}.
\end{multline}
This case can be called partially inhomogeneous model: the model is
homogeneous along one direction, but still inhomogeneous in the other one.

Having in mind this case as the starting point, one can
consider, for any chosen $s$ ($s=1,\dots,N$), the case
where $\lambda_1,\dots,\lambda_s$ are
different, but $\lambda_{s+1}=\dots=\lambda_{N}=\lambda$. One can still refer
to
such a situation as the partially inhomogeneous model.
It turns out that in this case, the partition function is closely related
to function $h_{N,s}(u_1,\dots,u_s)$ given by \eqref{hNs-def}; in particular,
the partition function \eqref{ZNpartial} is related to function
$h_{N,N}(u_1,\dots,u_N)$.

Let us define variables $\xi_1,\dots,\xi_N$ such that
\begin{equation}
\lambda_i=\lambda+\xi_i,\qquad i=1,\dots,N.
\end{equation}
It is convenient to introduce the `bare' partition function,
\begin{equation}
\wt Z_N(\xi_1,\xi_2,\dots,\xi_N):=
\frac{Z_N(\lambda_1,\lambda_2,\dots,\lambda_N)}{Z_N(\lambda,\lambda,\dots,\lambda)}.
\end{equation}
Below we shall treat $\xi_i$'s as variables while $\lambda$ is to be regarded
as
a parameter of the homogeneous model.

It is useful to consider first the case where all $\xi_i$'s are zeros but
one,
say $\xi_1$, is kept nonzero. Denoting $\xi:=\xi_1$ we straightforwardly have
\begin{equation}\label{wtZNxi}
\wt Z_N(\xi,0,\dots,0)=
\frac{(N-1)!}{(\sin\xi)^{N-1}}
\left[\frac{\varphi(\lambda)}{\varphi(\lambda+\xi)}\right]^{N}
\frac{\wt D_N(\xi)}{D_N}
\end{equation}
where, for later use, we have denoted
\begin{equation}\label{wtDN}
\wt D_N(\xi):=
\begin{vmatrix}
\varphi(\lambda)& \varphi'(\lambda)& \dots
& \varphi^{(N-2)}(\lambda) & \varphi(\lambda+\xi)\\
\varphi'(\lambda)& \varphi''(\lambda)& \dots & \varphi^{(N-1)}(\lambda)
& \varphi'(\lambda+\xi)\\
\vdots & \vdots & \ddots & \vdots &\vdots \\
\varphi^{(N-1)}(\lambda)& \varphi^{(N)}(\lambda)& \dots &
\varphi^{(2N-3)}(\lambda)
& \varphi^{(N-1)}(\lambda+\xi)
\end{vmatrix},
\end{equation}
and $D_N$ is given by \eqref{DN}. Note that in the case of
one nonzero inhomogeneity
one can always assume that it is attached to the last column
(since $Z_N(\lambda_1,\dots,\lambda_N)$ is a symmetric function
in its variables); thus we have chosen $\xi_2=\cdots=\xi_N=0$ above.

To proceed further let us
come back to the definition of the partition function as a sum over
all configurations.
The peculiarity of the domain wall boundary conditions is that they
admit only one vertex of weight $c$
in the last column; if this vertex is at $r$th position
(counted from the top) then the first
$(r-1)$ vertices are of weight $b$ while the remaining
$(N-r)$ vertices are of weight $a$. As it is has been explained in Section 3,
the probability of having the $c$-weight vertex at $r$th position on last
column is
equal to $H_N^{(r)}$, the correlation function which is originally defined as
the probability of having the $c$-weight vertex at $r$th position
on the first row, see \eqref{HNr-def}.
Due to this observation, the following relation is valid \cite{Ze-96},
\begin{equation}\label{sumr}
Z_N(\lambda_1,\lambda,\dots,\lambda)= Z_N
\sum_{r=1}^{N}
H_N^{(r)}
\left[\frac{\sin(\lambda_1-\eta)}{\sin(\lambda-\eta)}\right]^{r-1}
\left[\frac{\sin(\lambda_1+\eta)}{\sin(\lambda+\eta)}\right]^{N-r},
\end{equation}
where $Z_N=Z_N(\lambda,\dots,\lambda)$. Recalling that
$h_N(z)=\sum_{r=1}^{N}H_N^{(r)}z^{r-1}$
and denoting
\begin{equation}
\gamma(\xi)=
\frac{\sin(\lambda+\eta)}{\sin(\lambda-\eta)}\,
\frac{\sin(\lambda+\xi-\eta)}{\sin(\lambda+\xi+\eta)}
\end{equation}
we obtain that formula \eqref{sumr} actually
reads
\begin{equation}\label{wtZNhN}
\wt Z_N(\xi,0,\dots,0)=
\left[\frac{\sin(\lambda+\xi+\eta)}{\sin(\lambda+\eta)}\right]^{N-1}
h_N(\gamma(\xi)).
\end{equation}

More generally, as it has been first observed in \cite{CP-06} (at the special
case
$\lambda=\pi/2$) and proven in \cite{CP-07b}, for all $s=1,\dots,N$, one has
\begin{equation}\label{wtZNhNs}
\wt Z_N(\xi_1,\dots,\xi_s,0,\dots,0)=
\prod_{i=1}^{s}
\left[\frac{\sin(\lambda+\xi_i+\eta)}{\sin(\lambda+\eta)}\right]^{N-1}
h_{N,s} (u_1,\dots,u_s)\,,
\end{equation}
where $h_{N,s}(u_1,\dots,u_s)$ is defined in \eqref{hNs-def}, and
\begin{equation}\label{uj's}
u_i:=\gamma(\xi_i).
\end{equation}
The proof of \eqref{wtZNhNs} is based on
formula \eqref{ZNpartial} and some standard facts from the theory of
orthogonal
polynomials. We refer for further details and proofs to \cite{CP-07b}.

Finally we mention that formulae
\eqref{wtZNxi} and \eqref{wtZNhN} provide a representation for
generating function $h_N(z)$.
This representation is used in Appendix B for a derivation
of the large $N$ leading term of $\ln h_N(z)$.

\section{Large $N$ limit for function $h_N(z)$}

Our aim here is to show how formula \eqref{hNlargeN} for the large $N$
limit of function $h_N(z)$ can be derived. To this aim we extend here the
method of \cite{KZj-00}, where the leading term of $\ln Z_N$ have
been found. It is to be stressed that we assume that the weights
correspond to the disordered regime only (see \cite{Zj-00} for a
discussion of why the same approach cannot be used for the
anti-ferroelectric regime).

The approach is based on making use of some
differential equations which can be obtained by means of
the so-called Sylvester determinant identity.
This identity relates the determinant of a matrix
with a determinant of some other matrix whose entries are minors of the
original matrix.
Namely, let $A$ be an $n$-by-$n$ matrix, with entries $A_{j,k}$. Let us
consider
minors of degree $p$ ($1\leq p \leq n$) of this matrix,
\begin{equation}
A\args{i_1,i_2,\dots,i_p}{k_1,k_2\dots,k_p}:=
\begin{vmatrix}
A_{i_1,k_1} & A_{i_1,k_2}& \dots & A_{i_1,k_p}\\
A_{i_2,k_1} & A_{i_2,k_2}& \dots & A_{i_2,k_p}\\
\vdots & \vdots& \ddots & \vdots \\
A_{i_p,k_1} & A_{i_p,k_2}& \dots & A_{i_p,k_p}
\end{vmatrix},
\end{equation}
where $1\leq i_1<i_2<\dots <i_p\leq n$ and $1\leq k_1<k_2<\dots <k_p\leq n$.
Let us introduce matrix $B$ such that
\begin{equation}
B_{ik}:= A\args{1,2,\dots,p,p+i}{1,2,\dots,p,p+k}, \qquad i,k=1,\dots,n-p.
\end{equation}
Hence, entries of $B$ are minors of matrix $A$ of degree $(p+1)$.
The Sylvester identity reads
\begin{equation}
\det A= \left[A\args{1,2,\dots,p}{1,2,\dots,p}\right]^{-(n-p-1)} \det B.
\end{equation}
The standard proof of this identity is based on the Gauss algorithm, see,
e.g.,
monograph \cite{G-00} for further details. In all examples
below we put $n=N+1$ and $p=N-1$, i.e., $B$ will be a two-by-two matrix.

Let us consider first the partition function, $Z_N$, given by
\eqref{ZN-Hankel}. We start with mentioning that for arbitrary $N$ the
partition function $Z_N=Z_N(\lambda)$ satisfies
\begin{equation}\label{ZNprops}
Z_N(\lambda)=Z_N(\pi-\lambda),\qquad
Z_N(\lambda)\big|_{\lambda=\eta}=(\sin2\eta)^{N^2}.
\end{equation}
Here, the first relation reflects the crossing symmetry of the model
(since $b(\lambda)=a(\pi-\lambda)$), and the
second one follows from the fact that if $b=0$ (or $a=0$) then there is
exactly one
configuration contributing to the partition function.
As $N\to \infty$, one has \cite{KZj-00}
\begin{equation}\label{ZNlargeN}
Z_N=\exp\left(-N^2 f +O(N)\right),
\end{equation}
where $f$ is the free energy per site, the quantity of interest. To find $f$,
let us
consider the H\"ankel determinant in \eqref{ZN-Hankel},
$D_N$, given by \eqref{DN}. As $N\to \infty$, we have
\begin{equation}\label{DNlargeN}
D_N= \left[\prod_{n=1}^{N-1}(n!)\right]^2 \exp\left(N^2\phi+O(N)\right),
\end{equation}
where $\phi$ and $f$ are related by
\begin{equation}\label{Fvarphi}
-f=\ln\sin(\lambda+\eta)+\ln\sin(\lambda-\eta)+\phi.
\end{equation}

To compute $\phi$, let us consider $D_{N+1}$ and apply the Sylvester
identity,
with
$n=N+1$ and $p=N-1$, which gives
\begin{equation}\label{Toda}
D_{N+1}= \frac{1}{D_{N-1}}
\begin{vmatrix}
\partial_\lambda^2 D_N & \partial_\lambda D_N\\
\partial_\lambda D_N & D_N^{}
\end{vmatrix}.
\end{equation}
In writing \eqref{Toda} we have taken into account that the
first derivative of a H\"ankel determinant
changes the last row or the last column only, and the second derivative
changes
both the last row and the last column only. Relation \eqref{Toda} can be
rewritten as
\begin{equation}\label{Toda2}
\frac{D_{N+1}D_{N-1}}{D_N^2}=\partial_\lambda^2\ln D_N.
\end{equation}
{}From \eqref{DNlargeN} it follows that, as $N\to\infty$,
the leading terms in each side of \eqref{Toda2} are of order $N^2$; marching
terms gives:
\begin{equation}
\rme^{2\phi}=\partial_\lambda^2\phi.
\end{equation}
To solve this equation it is convenient to define $W:=\exp(-\phi)$,
so that we arrive at
\begin{equation}
(\partial_\lambda W)^2 -W\partial_\lambda^2 W=1.
\end{equation}
Obviously, the general solution of this equation has the form
\begin{equation}
W=\frac{\sin(\alpha_1\lambda+\alpha_2)}{\alpha_1},
\end{equation}
where $\alpha_1$ and $\alpha_2$ are some constants.
To fix these constants, let us turn to properties \eqref{ZNprops}, and via
relation
\eqref{Fvarphi} we obtain that they imply $W(\lambda)=W(\pi-\lambda)$ and
$W(\lambda)\big|_{\lambda=\eta}=0$, respectively. We find
\begin{equation}
\alpha_2=\frac{\pi}{2}(1-\alpha_1),\qquad
\alpha_1=\frac{\pi}{\pi-2\eta}.
\end{equation}
In terms of parameter $\alpha$ given by \eqref{alpha}, we have
$\alpha_1=\alpha$ and $\alpha_2=-\eta\alpha$, and therefore
the result for $\phi$ reads
\begin{equation}\label{phi}
\phi=\ln\left(\frac{\alpha}{\sin\alpha(\lambda-\eta)}\right).
\end{equation}
Substituting this result into \eqref{Fvarphi}, we readily reproduce
formula \eqref{freeenergy} for the free energy in the disordered regime.

A detailed analysis of the large $N$  expansion for the partition
function of the domain-wall six-vertex  model can be found in \cite{BF-05}.

Let us now turn to the function $h_N(z)$.
Due to formula \eqref{wtZNhN}, the large $N$ limit of function $h_N(z)$
can be found from that of the partition function $\wt Z_N(\xi,0,\dots,0)$.
This quantity is given by formula \eqref{wtZNxi}, where the nontrivial object
is the
last factor, the ratio of two determinants, which we denote
\begin{equation}\label{SN}
S_N(\xi):=\frac{\wt D_N(\xi)}{D_N}
\end{equation}
(note that all quantities here also depend on $\lambda$, which is to be
regarded now as a
parameter). For any finite $N$ we have the property
(recall that $\wt Z_N(\xi,0,\dots,0)|_{\xi=0}=1$)
\begin{equation}\label{xitozero}
S_N(\xi)\sim \frac{\xi^{N-1}}{(N-1)!},\qquad \xi\to 0.
\end{equation}
{}From the interpretation of $\wt Z_N(\xi,0,\dots,0)$ as the `bare' partition
function
it follows that it can grow up at most exponentially as $N$ increases,
so that, as $N\to\infty$, we have
\begin{equation}\label{SNlargeN}
S_N(\xi)=\frac{1}{(N-1)!}\exp(N\psi(\xi)+o(N)).
\end{equation}
We show now that $\psi(\xi)$ can be found as a solution
of an ordinary first order differential equation (in the variable $\xi$).
In turn, the result will imply formula \eqref{hNlargeN}.

Let us again use Sylvester identity, with $n=N+1$ and $p=N-1$,
applying it to $\wt D_{N+1}$ and to $\partial_\lambda \wt D_{N+1}$ (see
\eqref{wtDN}),
that gives
\begin{equation}
\begin{split}
\wt D_{N+1}(\xi)
& = \frac{1}{D_{N-1}}
\begin{vmatrix}
\partial_\lambda\wt D_N(\xi) & \partial_\lambda D_N \\
\wt D_N(\xi) & D_N
\end{vmatrix},
\\
\partial_\lambda \wt D_{N+1}(\xi)
& = \partial_\xi \wt D_{N+1}(\xi)
+\frac{1}{D_{N-1}}
\begin{vmatrix}
\partial_\lambda\wt D_N(\xi) & \partial_\lambda^2 D_N \\
\wt D_N(\xi) & \partial_\lambda D_N
\end{vmatrix}.
\end{split}
\end{equation}
In terms of function \eqref{SN} these two relations read:
\begin{equation}
\begin{split}
\frac{D_{N+1}D_{N-1}}{D_N^2}\, S_{N+1}(\xi)&=\partial_\lambda S_N(\xi)
\\
\partial_\lambda S_{N+1}(\xi)
+ \frac{\partial_\lambda D_{N+1}}{D_{N+1}}\, S_{N+1}(\xi)
&= \partial_\xi S_{N+1}(\xi)
+\frac{D_N\partial_\lambda D_N}{D_{N+1}D_{N-1}}\, \partial_\lambda S_N(\xi),
\\ &\qquad
+\frac{(\partial_\lambda D_N)^2-D_N \partial_\lambda^2 D_N}{D_{N+1}D_{N-1}}\,
S_N(\xi).
\end{split}
\end{equation}
The first relation here allows us to eliminate $\partial_\lambda S_N(\xi)$
and $\partial_\lambda S_{N+1}(\xi)$ in the second relation, and further,
using
\eqref{Toda2} and shifting $N\mapsto N-1$, we obtain
\begin{equation}\label{eqSN}
\frac{D_{N+1}D_{N-1}}{D_N^2}\, S_{N+1}(\xi)
+ \left(\partial_\lambda \ln \frac{D_N}{D_{N-1}}\right)
S_N(\xi) +S_{N-1}(\xi) = \partial_{\xi}S_N(\xi).
\end{equation}

After dividing by $S_N(\xi)$, both sides of relation \eqref{eqSN} are of
order $N$ in the large $N$ limit; matching terms leads to the following
ordinary first order differential equation:
\begin{equation}
\rme^{2\phi}\,\rme^{\psi(\xi)}+2\partial_\lambda\phi
+ \rme^{-\psi(\xi)}=\partial_\xi\psi(\xi).
\end{equation}
Here the function $\phi$ is given by \eqref{phi}; explicitly, the equation
reads
\begin{equation}
\frac{\alpha^2}{\sin\!^2\alpha(\lambda-\eta)}\, \rme^{\psi(\xi)}-
2\alpha\cot\alpha(\lambda-\eta)+\rme^{-\psi(\xi)}=\partial_\xi\psi(\xi).
\end{equation}
Due to \eqref{xitozero} we need the solution of this equation such
that $\exp\psi(\xi)\sim\xi$ when $\xi\to0$. The solution reads
\begin{equation}
\rme^{\psi(\xi)}=\frac{\sin\alpha(\lambda-\eta)\sin\alpha\xi}
{\alpha\sin\alpha(\xi+\lambda-\eta)}.
\end{equation}
Recalling formulae \eqref{wtZNxi}, \eqref{SN}, and \eqref{SNlargeN}, we
therefore find that, as $N\to\infty$,
\begin{equation}
\ln \wt Z_N(\xi,0,\dots,0)=N
\ln\left(
\frac{\varphi(\lambda)}{\varphi(\lambda+\xi)}
\, \frac{\sin\alpha(\lambda-\eta)\sin\alpha\xi}
{\alpha\sin\alpha(\xi+\lambda-\eta)\sin\xi}\right)+o(N).
\end{equation}
Finally, recalling that function $\varphi(\lambda)$ is given by
\eqref{varphi} and that function $h_N(z)$ is related to
$\wt Z_N(\xi,0,\dots,0)$ by \eqref{wtZNhN}, we readily obtain expression
\eqref{hNlargeN}, which is thus proven.

\bibliographystyle{amsplain}
\bibliography{condlong_bib}
\end{document}